\documentclass[useAMS,usenatbib]{mn2e}

\usepackage{graphicx}	
\usepackage{amsmath}	
\usepackage{amssymb}	
\usepackage{multicol}        
\usepackage{bm}		
\usepackage{pdflscape}	
\usepackage{multirow,multicol} 

\def\N1316{NGC\,1316}
\def\N1404{NGC\,1404}

\def\4U{4U~1735$-$444}

\def\arcsec{\ifmmode '' \else $''$\fi}

\def\arcsecpoint{\ifmmode ''\!. \else $''\!.$\fi}

\def\kms{\ifmmode {\rm km\ s}^{-1} \else km s$^{-1}$\fi}
\def\Msun{\ifmmode {\rm M}_{\odot} \else M$_{\odot}$\fi}
\def\Lsun{\ifmmode {\rm L}_{\odot} \else L$_{\odot}$\fi}
\def\Zsun{\ifmmode {\rm Z}_{\odot} \else Z$_{\odot}$\fi}

\def\ergscm2{ergs\,s$^{-1}$\,cm$^{-2}$}
\def\icm3{{\rm cm}^{-3}}
\def\icm2{{\rm cm}^{-2}}
\def\qo{\ifmmode q_{\rm o} \else $q_{\rm o}$\fi}
\def\Ho{\ifmmode H_{\rm o} \else $H_{\rm o}$\fi}
\def\ho{\ifmmode h_{\rm o} \else $h_{\rm o}$\fi}

\def\vFWHM{\ifmmode v_{\mbox{\tiny FWHM}} \else
            $v_{\mbox{\tiny FWHM}}$\fi}
\def\CCF{\ifmmode F_{\it CCF} \else $F_{\it CCF}$\fi}
\def\ACF{\ifmmode F_{\it ACF} \else $F_{\it ACF}$\fi}
\def\Halpha{\ifmmode {\rm H}\alpha \else H$\alpha$\fi}
\def\Hbeta{\ifmmode {\rm H}\beta \else H$\beta$\fi}
\def\Hgamma{\ifmmode {\rm H}\gamma \else H$\gamma$\fi}
\def\Hdelta{\ifmmode {\rm H}\delta \else H$\delta$\fi}
\def\Lya{\ifmmode {\rm Ly}\alpha \else Ly$\alpha$\fi}
\def\Lyb{\ifmmode {\rm Ly}\beta \else Ly$\beta$\fi}
\def\Lyg{\ifmmode {\rm Ly}\beta \else Ly$\gamma$\fi}

\def\ciii{\ifmmode {\rm C}\,{\sc iii} \else C\,{\sc iii}\fi}
\def\civ{\ifmmode {\rm C}\,{\sc iv} \else C\,{\sc iv}\fi}
\def\cv{\ifmmode {\rm C}\,{\sc v} \else C\,{\sc v}\fi}
\def\cvi{\ifmmode {\rm C}\,{\sc vi} \else C\,{\sc vi}\fi}

\def\o5007{[O\,{\sc iii}]\,$\lambda5007$}

\def\ovii{O\,{\sc vii}}
\def\oviii{O\,{\sc viii}}

\def\nex{Ne\,{\sc x}}

\def\fexvii{Fe\,{\sc xvii}}

\def\fexxii-iii{Fe\,{\sc xxii-xxiii}}

\usepackage{xcolor}

\usepackage{hyperref}

\hypersetup{colorlinks=true,linkcolor=blue,citecolor=blue,filecolor=blue,urlcolor=blue}

\title[Wind vs. state variability in ULX NGC 1313 X-1]
{\centering{XMM-\textit{Newton} Campaign On Ultraluminous X-ray Source NGC 1313 X-1: Wind vs. State Variability}}
\author[C. Pinto et al.]{C. Pinto,$^{1,2,3}$\thanks{E-mail:
ciro.pinto@esa.int} D. J. Walton,$^{2}$ E. Kara,$^{4}$ M. L. Parker,$^{5}$ R. Soria,$^{6,7}$ P. Kosec,$^{2}$ 
\newauthor M. J. Middleton,$^{8}$ W. N. Alston,$^{2}$ A. C. Fabian,$^{2}$ M. Guainazzi,$^{1}$ T. P. Roberts,$^{9}$ 
\newauthor F. Fuerst,$^{5}$ H. P. Earnshaw,$^{10}$ R. Sathyaprakash\,$^{9}$ and D. Barret\,$^{11}$ \\
$^{1}$ESTEC/ESA, Keplerlaan 1, 2201AZ Noordwijk, The Netherlands\\
$^{2}$Institute of Astronomy, Madingley Road, CB3 0HA Cambridge, United Kingdom\\
$^{3}$INAF - IASF Palermo, Via U. La Malfa 153, I-90146 Palermo, Italy\\
$^{4}$Massachusetts Institute of Technology, Kavli Institute, Cambridge, MA, USA\\
$^{5}$ESAC/ESA European Space Astronomy Center, P.O. Box 78, 28691 Villanueva de la Canada, Madrid, Spain\\
$^{6}$College of Astronomy and Space Sciences, University of the Chinese Academy of Sciences, Beijing 100049, China\\
$^{7}$Sydney Institute for Astronomy, School of Physics A28, The University of Sydney, Sydney, NSW 2006, Australia\\
$^{8}$Physics \& Astronomy, University of Southampton, Southampton, Hampshire SO17 1BJ, UK\\
$^{9}$Centre for Extragalactic Astronomy, Durham University, Department of Physics, South Road, Durham DH1 3LE, UK\\
$^{10}$Cahill Center for Astronomy and Astrophysics, California Institute of Technology, Pasadena, CA 91125, USA\\
$^{11}$Universit\'e de Toulouse, CNRS, IRAP, 9 Avenue du colonel Roche, BP 44346, 31028 Toulouse Cedex 4, France}
\begin{document}

\date{\today}

\pagerange{\pageref{firstpage}--\pageref{lastpage}} \pubyear{2018}

\maketitle

\label{firstpage}

\begin{abstract}
Most ultraluminous X-ray sources (ULXs) are thought to be powered by 
neutron stars and black holes accreting beyond the Eddington limit.
\textcolor{black}{If the compact object is a black hole 
or a neutron star with a magnetic field $\lesssim10^{12} G$},
the accretion disc is expected to thicken and launch powerful winds driven by radiation pressure.
Evidence of such winds has been found in ULXs through the high-resolution
spectrometers onboard XMM-\textit{Newton}, but several unknowns remain, 
such as the geometry and launching mechanism of these winds.
In order to better understand ULX winds and their link 
to the accretion regime, we have undertaken a major campaign with XMM-\textit{Newton}
to study the ULX NGC 1313 X-1, which is known to exhibit strong emission and absorption
features from a mildly-relativistic wind.
The new observations show clear changes in the wind 
with a significantly weakened fast component ($0.2c$)
and the rise of a new wind phase which is cooler and slower (0.06$-$0.08$c$).
We also \textcolor{black}{detect for the first time variability in the emission lines} which indicates an origin 
within the accretion disc or in the wind.
We describe the variability of the wind 
in the framework of variable super-Eddington accretion rate
and discuss a possible geometry for the accretion disc.
\end{abstract}


\section{Introduction}
\label{sec:intro}

Ultraluminous X-ray sources (ULXs) are bright, point-like, off-nucleus,
extragalactic sources with X-ray luminosities above $10^{39}$ erg/s
that result from accretion onto a compact object (see, e.g., \citealt{Kaaret2017}). 
Recent studies have shown that some ULXs are powered by
accretion onto neutron stars with strong magnetic fields 
(\textcolor{black}{$10^{12-14} G$}, e.g. \citealt{Bachetti2014}, 
\citealt{Fuerst2016}, \citealt{Israel2017a}, \citealt{Israel2017b},
\citealt{Tsygankov2016}, \citealt{Carpano2018}, \citealt{Brightman2018}, 
\citealt{RodriguezCastillo2019}, \citealt{Sathyaprakash2019a},
\citealt{Middleton2019}).
Moreover, since more than a decade it has been thought that the majority of ULXs 
consists of stellar-mass compact objects $(<100 M_{\odot})$ at or in
excess of the Eddington limit (\citealt{King2001, Poutanen2007, 
Gladstone2009, Middleton2013, Liu2013, Motch2014}). 
In particular, the presence of a strong turnover below $\sim7$\,keV 
in most ULX \textcolor{black}{(and all \textit{NuSTAR})} spectra 
(see, e.g., \citealt{Gladstone2009}, \citealt{Bachetti2013} and
\citealt{Walton2018a})
and the simultaneous high $10^{39-41}$\,erg/s luminosities are difficult to explain 
with sub-Eddington accretion models.

{However, some ULXs might still host more massive black holes
($10^{2-5} M_{\odot}$) at more sedate Eddington ratios 
(see, e.g., \citealt{Greene2007, Farrell2009, Webb2012, Mezcua2016}).}

At accretion rates approaching the Eddington limit,
radiation pressure inflates the thin disc,
producing a geometry similar to that of a funnel and
a wind is launched (see, e.g., \citealt{SS1973} and \citealt{Poutanen2007}). 
Winds driven by radiation pressure in a super-Eddington (or supercritical) regime 
can reach mildly-relativistic velocities ($\sim0.1c$, see \citealt{Takeuchi2013}) 
and therefore may carry a huge amount of matter and kinetic energy out of the system.
They can affect the accretion onto the compact object and 
the physical state of the nearby interstellar medium by inflating large
($\sim100$ pc) bubbles. For a quantitative comparison between the wind power
and the energetics of the bubbles see \citet{Pinto2019a} 
and references therein.


Evidence for such winds was first seen through strong residuals
at soft X-ray energies ($<$\,2\,keV) to the 
best-fitting models describing the continuum of ULXs, albeit unresolved,
in the low-to-moderate resolution CCD spectra (see, e.g., \citealt{Stobbart2006}). 
The time variability of such residuals and its correlation with the source spectral
hardness suggested that they might be produced by the ULX itself and possibly in the 
form of a wind (see, e.g., \citealt{Middleton2014} and \citealt{Middleton2015b}).

Finally, our team \citep{Pinto2016nature} identified strong ($>5\sigma$) 
rest-frame emission and blue-shifted ($\sim0.2c$) absorption lines in the 
high-resolution X-ray spectra of two bright ULXs, namely NGC 1313 X-1
and NGC 5408 X-1, taken with the XMM-\textit{Newton} 
Reflection Grating spectrometer (RGS). 
\citet{Walton2016a} found an Fe\,K component to the outflow 
in NGC 1313 X-1 significant at the $3\sigma$ level
through the combination of hard X-ray spectra taken with XMM-\textit{Newton} 
and \textit{NuSTAR} (the \textit{Nuclear Spectroscopic Telescope Array}, 
\citealt{Harrison2013}).
Additional evidence of powerful winds was found in the RGS spectra of 
other sources such as the soft NGC 55 ULX \citep{Pinto2017} and the intermediate hardness
NGC 5204 ULX-1 \citep{Kosec2018a}. 
\textcolor{black}{Recently, \citet{Wang2019} found evidence of blueshifted emission lines 
in \textit{Chandra} and XMM-\textit{Newton} spectra of a transient ULX in NGC 6946}. 
The first discovery of outflows in
the ultraluminous X-ray pulsar NGC 300 PULX
\citep{Kosec2018b} showed that strong magnetic fields (e.g $\sim10^{12} G$) and 
winds can coexist as is also shown by the theoretical calculation of \citet{Mushtukov2019a}.

In the last decade, a unification scenario for ULXs has gained popularity, which 
proposes that the observed behaviour from ULXs (i.e. broadband spectra, variability properties) 
depends on both their accretion rates and viewing angles
(see, e.g., \citealt{Poutanen2007, Middleton2015a, Feng2016, Urquhart2016},
\citealt{Pinto2017}).
\textcolor{black}{Moreover, we should expect higher ionisation states in the wind at low
inclinations (on-axis) due to exposure to harder SEDs in the inner regions of the disc (\citealt{Pinto2019a}).}
The study of the winds provides an alternative and complementary
tool to understand the physics of ULXs and super-Eddington accretion 
to the modelling of the broadband spectral continuum.


Despite important discoveries and the progress on ULXs in the last decade, 
several problems are still to be solved:
How does the wind vary with the accretion rate?
What is the geometry and the ionisation structure of the wind? 
What is the launching mechanism? What is its energy budget?
Can the wind affect the ULX surroundings and the accretion of matter onto the 
compact object? If the compact object is a magnetised neutron star, could 
a strong magnetosphere disrupt the disc and suppress the wind?
\textcolor{black}{When and how jets are launched in ULXs?}


In order to place some constraints on these unknowns we were awarded 
6 orbits targeting the NGC1313 galaxy 
with XMM-\textit{Newton} during AO16 (PI: Pinto) and obtained quasi-simultaneous 
\textit{NuSTAR} coverage of the hard X-ray band (PI: Walton) for some observations. 
Additional observations were taken with \textit{Chandra} (PI: Canizares).
This is the first paper in a series focusing on NCG1313 X-1, and presents a study of 
its wind as seen by XMM-\textit{Newton} and, in particular, its variability with the ULX spectral state.
Detailed broadband spectroscopy and the analysis of the \textit{NuSTAR} data
is performed in \citet{Walton2020},
while the search for narrow spectral lines in the hard band with the \textit{Chandra}
gratings (HETGS) will be shown in Nowak et al. (in prep).
Our campaign has already provided important results such as the first
discovery of pulsations in the NGC 1313 X-2 \citep{Sathyaprakash2019a},
which added a new PULX candidate to a list which includes 
only a handful of sources at the present.
\textcolor{black}{We also report the discovery of a soft X-ray lag in new XMM-\textit{Newton} 
data of NGC 1313 X-1 from this campaign in a companion paper \citep{Kara2020}.
Soft lags were discovered only in two other ULXs.}

We detail our campaign in Sect.\,\ref{sec:data} and show
the spectral analysis in Sect.\,\ref{sec:spectral_analysis}. 
In Sect.\,\ref{sec:timing_analysis} we show the principal component analysis.
We discuss the results in Sect.\,\ref{sec:discussion} 
and give our conclusions in Sect.\,\ref{sec:conclusion}.
Throughout the paper we use C-statistics \citep{Cash1979}, adopt 1\,$\sigma$ error bars
and perform the spectral analysis with the {\scriptsize{SPEX}} code{\footnote{http://www.sron.nl/spex}}
\citep{kaastraspex}.

\begin{table*}
\caption{XMM-\textit{Newton} observations used in this paper.}  
\label{table:obs_log}     
\renewcommand{\arraystretch}{1.1}
 \small\addtolength{\tabcolsep}{+2pt}
 
\scalebox{1}{%
\begin{tabular}{c c c c c c c c c c c c c c c c c c}     
\hline  
OBS\_ID  &  Date & State & Flux  & t$_{\rm RGS1,2}$ & t$_{\rm pn}$ & t$_{\rm MOS1}$ & t$_{\rm MOS2}$ & Notes & Stacking \\
                         &               &            & $\rm{photons \, m}^{-2}\,\rm{s}^{-1}$    &  (ks)  & (ks)   & (ks) & (ks) &      & Group \\
\hline                                                                                                                       
 0405090101    &  2006-10-15  &  Low-Int      &   7.8   &    97    &  81    &  99     &  99      &  &  OLD L-I \\
 0693850501    &  2012-12-16  &  Low-Int      &  10.1   &     112  &   90  &   114  &   113   &  &  OLD L-I  \\
 0693851201    &  2012-12-22  &  Low-Int      &  11.0  &    117   &  85   &  122   &  121      &  & OLD L-I  \\
\hline                                                                                                                                                             
 0742590301    &  2014-07-05  &  Int-Bright   &  18.1  &    62    &  54   &  61       &  61    &  $^{(*)}$ \\
 0794580601    &  2017-03-29  &  Low-Int      &  11.0  &    41    &  26   &  39       &  37    &  $^{(*)}$ \\
\hline                                                                                                  
 0803990101    &  2017-06-14  &  Int-Bright  & 15.5   &    131  &  109   &  128   &  128    &   &  NEW I-B \\
 0803990201    &  2017-06-20  &  Int-Bright  & 13.5   &    130  &  111   &  129   &  128   &    &  NEW I-B \\
 0803990301    &  2017-08-31  &  Low-Int     &  8.6     &    109 &  42    &  70      &  64      & $^{(**)}$    &  NEW L-I  \\
 0803990401    &  2017-09-02  &  Low-Int     &  8.5     &    115  &  53    &  50      &  46     &  $^{(**)}$   &  NEW L-I    \\
 0803990701    &  2017-09-24  &  Low-Int     &  9.1    &    121  &  11    &  10       &  10     &   $^{(***)}$  & NEW L-I    \\
 0803990501    &  2017-12-07  &  Int-Bright  & 12.0    &    90   &  60     &  89      &  86     &  $^{(**)}$  &  NEW I-B \\
 0803990601    &  2017-12-09  &  Int-Bright  & 14.7    &    109 &  71    &  117    &  116   &  $^{(**)}$   &  NEW I-B \\
\hline                
\end{tabular}}

Notes: the ULX \textit{state} refers to the low-to-intermediate and intermediate-to-bright 
flux levels, primarily chosen from the total flux and shape of the X-ray spectra 
(see Fig.\,\ref{Fig:Plot_Swift_all} and \ref{Fig:Plot_EPIC_all}). 
Absorbed 0.3$-$10 keV fluxes are computed with a multi-component model consisting
of two modified blackbody components plus comptonisation corrected for 
Galactic absorption (see Sect.\,\ref{sec:baseline_continuum}). Exposure times account for background flaring removal. 
$^{(*)}$\,RGS spectra contaminated by either X-2 or SNR (only EPIC-pn is used for principal component analysis);
$^{(**)}$\,EPIC cameras affected by solar flares; 
$^{(***)}$\,Additional calibration exposure (only RGS is operating for the full orbit).
{The acronyms OLD L-I, NEW L-I  and NEW I-B stay for old low-intermediate, new low-intermediate
and new intermediate-bright states.}
\end{table*}

\begin{table}
\caption{\textit{NuSTAR} observations used in this paper.}  
\centering
\label{table:obs_log_nustar}   
\renewcommand{\arraystretch}{1.1}
 \small\addtolength{\tabcolsep}{+2pt}
 
\scalebox{1}{%
\begin{tabular}{c c c c c c c c c c c c c c c c c c}     
\hline  
OBS\_ID  &  Date   & t$_{\rm FMPA,B}$  & Notes \\
                &             &   (ks)                      &            \\
\hline                                                                                                                       
 30002035002    &  2012-12-16     &    154     &  $^{(*)}$   \\
 30002035004    &  2012-12-21     &    206     &  $^{(*)}$  \\
\hline                                                                             
 80001032002    &  2014-07-05     &    73       &  $^{(*)}$ \\
 90201050002    &  2017-03-29     &    125     &  $^{(*)}$ \\
\hline                                                             
 30302016002    &  2017-06-14     &    100    &  $^{(**)}$    \\
 30302016004    &  2017-07-17     &    87      &  $^{(***)}$   \\
 30302016006    &  2017-09-03     &    90      &  $^{(**)}$     \\
 30302016008    &  2017-09-15     &    108    &  $^{(***)}$   \\
 30302016010    &  2017-12-09     &    100    &  $^{(**)}$    \\
\hline                
\end{tabular}}

Notes: $^{(*,**)}$\,archival and new observations simultaneous with XMM-\textit{Newton},
           $^{(***)}$\,new observations simultaneous with \textit{Chandra}.
            All exposures are given to the nearest ks. 
\end{table}


\section[]{The data products}
\label{sec:data}

\subsection[]{NGC 1313 X-1 archival data}

NGC 1313 X-1 is an archetypal bright ULX ($L_{\,0.3-10\,{\rm keV}}$ up to $\sim10^{40}$ erg/s).
Its relatively small distance ($D=4.2$\,Mpc, e.g. \citealt{Tully2013}) makes it a good target 
for the high-spectral-resolution 
Reflection Grating Spectrometers (RGS) aboard XMM-\textit{Newton}, 
which cover the 0.3$-$2\,keV energy band where strong (unresolved) features 
were found in low-spectral resolution CCD spectra \citep{Middleton2015b}.
The strength of these features decreases at higher \textcolor{black}{luminosity} and spectral hardness
(they define as hardness the ratio of the unabsorbed 1$-$10 / 0.3$-$1 keV fluxes, 
\textcolor{black}{while in this paper we use the standard 2$-$10 / 0.3$-$2 keV flux ratio,
so our brightest states are also the softest}). 
The RGS spectra resolve the spectral features in a forest of rest-frame emission 
lines and relativistically-blueshifted ($\sim0.2c$) absorption lines from a wind
as shown by \citet{Pinto2016nature}. In that paper, high-spatial resolution \textit{Chandra} 
maps indicate that the X-ray emission is confined to a circular region 
of a few arcseconds, confirming that the residuals are intrinsic to the ULX
as shown by \citet{Sutton2015} and \citet{Sathyaprakash2019b} 
for NGC 5408 X-1 and Holmberg IX X-1.

XMM-\textit{Newton} has observed the NGC 1313 galaxy before our campaign for 25 times
with exposures ranging from 9\,ks (snapshot) to 137\,ks (full orbit). However,
of these observations, only three had ULX-1 on-axis for a significant
amount of time and with an appropriate roll angle. 

\citet{Kosec2018a} have shown that observations of
about 100 ks are normally required to detect and resolve strong lines 
in ULX RGS spectra.
Moreover, RGS observations have to be performed with accurate pointing (on-axis) and roll angle
in order to avoid contamination from any other bright source located along the (long) 
dispersion direction axis of the RGS field of view ($\sim5'\times30'$).

In Table\,\ref{table:obs_log} we provide some essential information on the deepest
observations of NGC 1313 X-1 (on-axis) taken with XMM-\textit{Newton}. 
The three old observations (October 2006 and December 2012) are the same used in 
\citet{Pinto2016nature} and \citet{Walton2016a}. 
The two following observations (July 2014, March 2017)
provide a sufficient amount of exposure time, but the roll angle is not optimal since 
the RGS selection region for ULX-1 intercept either ULX-2 or the supernova remnant  
SN 1978K. These two observations are therefore not used in our 
high-spectral-resolution analysis.

\subsection[]{NGC 1313 X-1 XMM-\textit{Newton} campaign}
\label{sec:xmm_campaign}

\begin{figure}
  \includegraphics[width=1\columnwidth, angle=0]{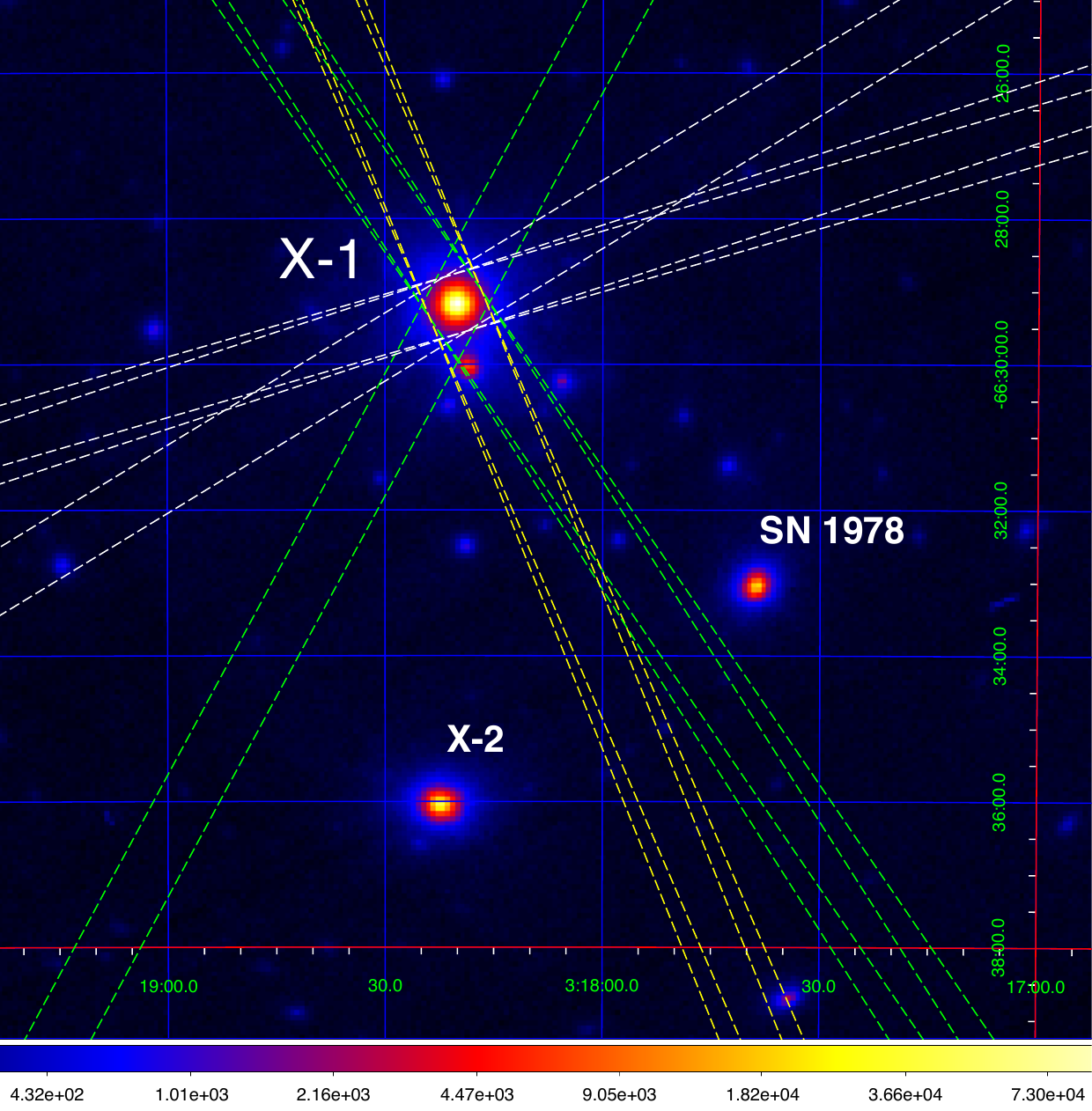}
   \caption{XMM-\textit{Newton} image of the NGC 1313 field
   obtained by combining all the data available
   from EPIC-pn and MOS\,1,2. The green, yellow and white dashed lines indicate
   the RGS extraction region for the old Low-Int, new Low-Int and new Int-Bright spectra,
   respectively \textcolor{black}{(see notation in Table\,\ref{table:obs_log})}. 
   The other ULX (X-2) and the supernova remnant (SN 1978)
   are also labelled. {The apparent lower number of datasets is due to the fact
   that some pointings have the same average position angles.} }
   \label{Fig:Plot_EPIC_image}
\end{figure}

We observed the NGC 1313 galaxy in 2017 
with a deeper campaign led by XMM-\textit{Newton}.
The roll angle was chosen to avoid any contamination along the dispersed RGS spectra
from the two nearby brightest X-ray sources 
(ULX-2 and SN 1978K, see Fig.\,\ref{Fig:Plot_EPIC_image}).
In addition, we observed the source in three different epochs separated by 2-3 months during 2017.
The primary goals were a) to detect the source in different spectral states
and b) to study the variability of the wind according to the spectral state.

The choice of the time windows was dictated by the historical lightcurve of 
NGC 1313 X-1 taken with the \textit{Neil Gehrels Swift Observatory} 
(hereafter \textit{Swift}; \citealt{Gehrels2004}), 
which shows long-term variability on timescales of 2-3 months
(see Fig.\,\ref{Fig:Plot_Swift_all}). Such variability is of crucial importance as
a $\sim$\,60 days modulation has been detected in some ULXs and interpreted
as the orbital or the super-orbital period of the binary system 
(see, e.g., \citealt{Kaaret2006, Motch2014, Walton2016b, Fuerst2018, Brightman2019}).
Observations of NGC 1313 X-1 separated by 2-3 months may therefore 
provide insights on the geometry of the ULX system
and the complex structure of the wind.
More detail on the analysis of the \textit{Swift} observations and the long
term variability of NGC 1313 X-1 will provided elsewhere 
(Middleton et al. in prep).

The observations of NGC 1313 X-1 were therefore divided in three groups
each consisting of a pair of full orbits within a few days of each other in
order to obtain a combined, deep ($\gtrsim$\,200 ks), RGS spectrum of the ULX 
representative of a specific flux regime or accretion rate.

In addition to the 6 orbits awarded during AO16, one additional orbit was awarded 
to perform EPIC calibration.
During this observation, only RGS and OM could be operated for the full 
duration of the observation, while the CCD cameras (pn and MOS 1,2)
were active for only approximately 10 ks each.
{We also use these observations, but their 
CCD data have limited statistical weight.}

\subsection[]{\textit{NuSTAR} and \textit{Chandra} follow-up}
\label{sec:nustar_campaign}

Details of the \textit{NuSTAR} archival and follow up observations are given 
in Table\,\ref{table:obs_log_nustar}. We also report whether the observations 
were simultaneous to XMM-\textit{Newton} or \textit{Chandra}.
The \textit{Chandra} observations taken throughout 2017 accumulate up to 500 ks 
of exposure time, but each is relatively short (between 10 and 50 ks).
\textcolor{black}{In this paper, we only use the \textit{NuSTAR} time-averaged spectrum to cover the high-energy
($>$ 10 keV) range to build up our broadband spectral energy distribution and perform accurate photoionisation modelling.}

\subsection[]{Data reduction}
\label{sec:data_reduction}

In this work we use data from the broadband European Photon-Imaging Camera 
(EPIC, \citealt{Turner2001}) and the Reflection Grating Spectrometer (RGS,
\citealt{denherder2001}) aboard XMM-\textit{Newton} 
 as well as the two focal plane modules (FPMA/B) aboard
\textit{NuSTAR} (\citealt{Harrison2013}).

The primary science is carried out 
with the RGS which can detect and resolve narrow 
spectral features. The broadband cameras (EPIC and FPMA/B) are mainly
used to provide the correct spectral continuum and cover the hard X-ray band
which is not covered by the RGS. The EPIC count rate is much higher than RGS
in the soft ($<\,2$ keV) band. 
However, the energy resolution of the EPIC CCD cameras is insufficient to resolve
narrow features, particularly in the soft band 
($R_{\,0.3-2 \, \rm keV, \, pn}=\Delta E/E \sim 10-20$). Therefore, we only use the EPIC-pn spectrum between 2 and 10 keV. 
Below 2 keV we only use RGS data 
($R_{\,0.3-2 \, \rm keV, \, RGS}=\Delta E/E \sim 100-600$
for the first order spectra).
We have decided not to use the MOS 1,2 cameras for spectroscopy 
\textcolor{black}{but just for imaging}
as each has 3$-$4 times less effective area (and even less above 8\,keV) than pn.
\textcolor{black}{Moreover, we mainly use broadband spectra to constrain the 
continuum shape and minimise its effects onto the RGS line search.}

The \textit{NuSTAR} data are only used here between 10 and 20 keV to cover the hard X-ray band
and determine the source continuum. This choice is dictated by the fact that 1) only a few 
\textit{NuSTAR} and XMM-\textit{Newton} observations are fully simultaneous and 
2) we want to maximise the \textcolor{black}{statistical weight} of the high-resolution RGS spectra. 
A detailed and in-depth analysis of the \textit{NuSTAR} 
data is presented in \citet{Walton2020}. Here we briefly give the main steps of the data 
reduction and the production of the time-averaged \textit{NuSTAR} spectrum.
We reduce the \textit{NuSTAR} data with the \textit{NuSTAR} Data Analysis 
Software ({\scriptsize{NuSTARDAS}} v1.8.0 and caldb v20171204). 
The unfiltered event files are cleaned with {\scriptsize{NUPIPELINE}}. 
We adopt the standard depth correction, which significantly reduces the internal background 
at high energies, and periods of earth-occultation and passages through the South Atlantic 
Anomaly are also excluded. Source products are extracted from circular regions of radius 30''
and the background is estimated from a larger square in a nearby region of the same detector free of 
contaminating point sources. Spectra and lightcurves are extracted from the cleaned event files 
using  {\scriptsize{NUPRODUCTS}} for both focal plane modules (FPMA and FPMB). 
In addition to the standard `science' data (mode 1), we also extract the `spacecraft science' data 
(mode 6) following \citet{Walton2016c}. 
Finally, all \textit{NuSTAR} spectra are combined into a single time-averaged spectrum 
using the {\scriptsize{FTOOL}}  {\scriptsize{ADDASCASPEC}}.


We reduce the 12 XMM-\textit{Newton} observations with the 
SAS\footnote{https://www.cosmos.esa.int/web/xmm-newton/sas} v16.0.0 
(CALDB available on February, 2019).
Briefly, EPIC-pn, MOS and RGS data are reduced with the \textit{epproc}, 
\textit{emproc} and \textit{rgsproc} tasks, respectively, 
to produce calibrated event files, spectra response matrices and images.
Following the standard procedures,
we filter the MOS and pn event lists for bad pixels, bad columns, cosmic-ray events 
outside the field of view (FOV), photons in the gaps (FLAG$=$0), 
and apply standard grade selection, corresponding to PATTERN $\leq12$
for MOS and PATTERN $\leq4$ for pn.
We correct for contamination from soft-proton flares through the SAS task 
\textit{evselect} by selecting background-quiescent intervals
on the light curves for MOS\,1,2 and pn in the 10$-$12 keV
energy band, while we use the data from the CCD number 9 for RGS
(corresponding to $\lesssim7.5$\,{\AA} or $\gtrsim1.7$\,keV). 
The light curves are grouped in 100\,s intervals and all the time bins 
with a count rate above 0.4 c/s are rejected for pn, 0.2 c/s for both MOS 
and RGS. 
We build the good time interval (GTI) files with the accepted time 
events for the pn, MOS and RGS data through the SAS task \textit{tabgtigen} and 
\textcolor{black}{filter the data with these GTI files}. 
For RGS 1 and 2 we join the GTI and obtain the same exposure times.
The RGS\,1-2, MOS\,1-2 and pn total clean exposure times are 
quoted in Table\,\ref{table:obs_log}.

\begin{figure*}
  \includegraphics[width=1.3\columnwidth, angle=0]{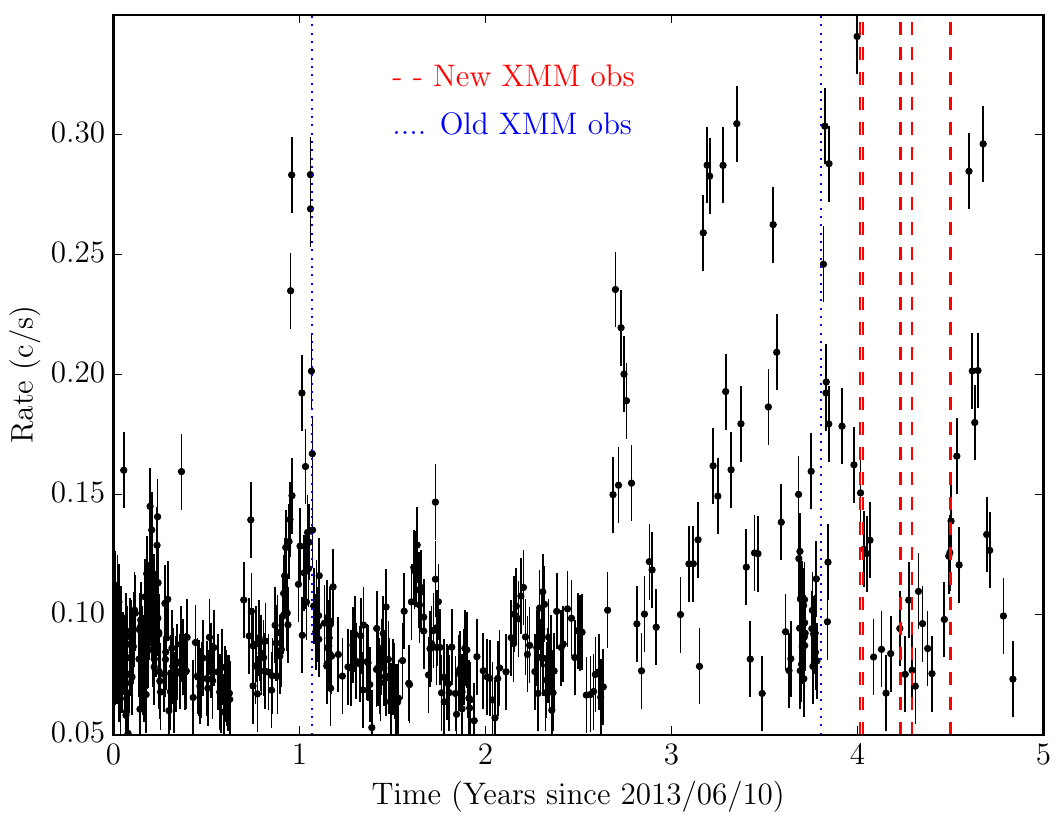}
  \includegraphics[width=0.69\columnwidth, angle=0]{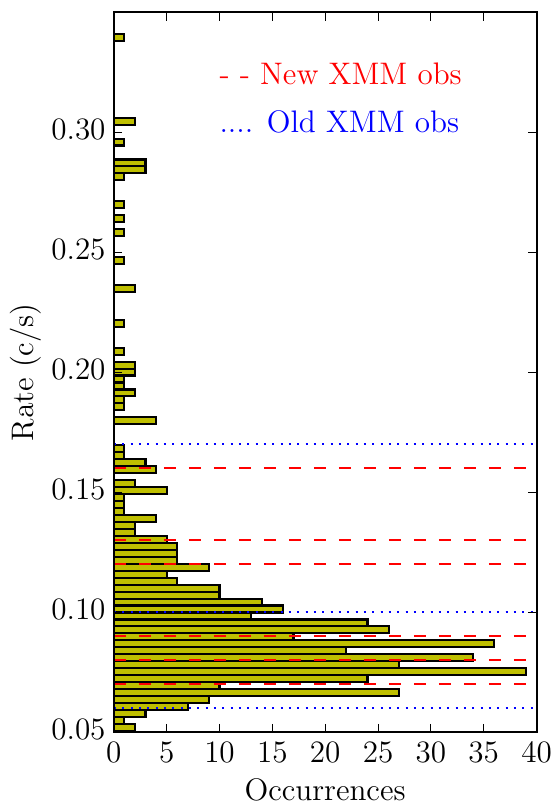}
   \vspace{-0.1cm}
   \caption{Left: Long-term \textit{Swift}/XRT lightcurve of NGC 1313 X-1 
   with the dates of the XMM-\textit{Newton} observations indicated by 
   dotted (old) and dashed (new, 2017) lines. \textcolor{black}{Right: Corresponding count rate histogram 
   with the XMM-\textit{Newton} observations indicated.}}
   \label{Fig:Plot_Swift_all}
   \vspace{-0.1cm}
\end{figure*}

Solar flares significantly affected the EPIC data throughout
the low-flux observations of our campaign for about 50\% of the exposure time.
Fortunately, solar flares affect the RGS data on a much lower level.
We extract EPIC MOS 1-2 and pn images in the 0.3$-$10\,keV energy range
and stack them with the \textit{emosaic} task
(see Fig.\,\ref{Fig:Plot_EPIC_image}).
We also extract EPIC-pn spectra from within a circular region of 30'' 
radius centred on the emission peak. 
The background spectra are extracted from within a larger box 
in a nearby region on the same chip, 
but away from bright sources and the readout direction
(see Fig.\,\ref{Fig:Plot_EPIC_image}).
We also make sure that the background region is not placed in
the copper emission region (\citealt{Lumb2002}).

We extract the first-order RGS spectra in a cross-dispersion region 
of 0.8' width, centred on the emission peak 
and the background spectra by selecting photons
beyond the 98\% of the source point-spread-function, making sure that
the background regions do not overlap with either X-2 or SN 1978, 
and check for consistency with blank field data. 

\subsubsection{Spectral states of NGC 1313 X-1}

The EPIC-pn spectra of NGC 1313 X-1 are shown in Fig.\,\ref{Fig:Plot_EPIC_all}.
It is clear that our observing strategy was appropriate because it 
enabled us to recovered a broad and continuous range of spectra ranging from the low flux
to very bright states (or hard ultraluminous and broadened disc,
e.g. \citealt{Sutton2013}). 
The archival observations provide further useful information and fill some 
small flux gaps in our new exposures (Fig.\,\ref{Fig:Plot_EPIC_all}).
It is not strictly accurate to speak of individual spectral states
as the source does not jump from one to another but rather moves 
smoothly among them. However, in order to simplify the terminology,
we still refer to low-to-intermediate (hard) and
intermediate-to-bright (broadened-disc) states (see also Table\,\ref{table:obs_log}).
\textcolor{black}{For instance, the other ULX in the same galaxy, NGC 1313 X-2,
seems to switch between two preferred states (see e.g. \citealt{Weng2018}), perhaps 
due to some processes related to the presence of strong magnetic fields
from the neutron star recently discovered \citep{Sathyaprakash2019a}.}

In order to better understand our spectra and place them within the context
of the source variability, we extract the historical lightcurve of NGC 1313 X-1
as obtained by \textit{Swift}/XRT using the website 
tool\footnote{http://www.swift.ac.uk/user\_objects/} \citep{Evans2009}
which processes the data
with {\scriptsize{HEASOFT}} v6.22. 
In Fig.\,\ref{Fig:Plot_Swift_all} (left) we show the \textit{Swift}/XRT lightcurve
from June 2013 where good sampling with intervals of a few days was available.
The source has shown a remarkable activity in the last 2 years
if compared to previous periods of time (more detail on the long term behaviour 
will be provided by Middleton et al. in prep).
The XMM-\textit{Newton} observations were taken
towards the end of a bright period (middle June 2017), during
the central part of its low-flux time range (early September) and finally in the beginning of the 
new rise (early December). 
Fig.\,\ref{Fig:Plot_Swift_all} (right) shows the histogram \textcolor{black}{of the countrate taken from} the
\textit{Swift}/XRT lightcurve with the dashed-red and dotted-blue lines
indicating the flux levels of the new and archival XMM-\textit{Newton} observations, respectively.
These flux levels are interpolated between the two closest \textit{Swift} points
in the lightcurve and provide 
rough approximations of the actual levels, useful to place the XMM-\textit{Newton} observations
in the general context of long-term source variability.

\subsubsection{Flux--/time--resolved spectra}
\label{sec:fluxed_spectra}

We stack the first-order RGS 1 and 2 spectra from the observations 
with \textcolor{black}{a similar flux level and} spectral state (either archival/old low-to-intermediate,
or new low-to-intermediate or new intermediate-to-bright) with the task
\textit{rgscombine} (see Table\,\ref{table:obs_log}). 
Similarly, we stack the EPIC-pn spectra for the archival and the new
observations using \textit{epicspeccombine}.
The \textit{NuSTAR} data are used mainly to cover the hard band 
and, since there is no major variability above 10 keV during our campaign
and in the archival spectra used in this paper \citep{Walton2020},
we prefer to stack all \textit{NuSTAR} FPMA,B spectra into a single deep spectrum 
and use only the data points between 10 and 20 keV.
We remind the reader that the wind search is driven by 
the high-resolution RGS spectra.
In summary, for the spectral analysis in this paper, we will focus on three pairs 
of simultaneous XMM-\textit{Newton}/RGS (0.4-2 keV) and EPIC-pn (2-10 keV) 
deep spectra and one \textit{NuSTAR} (10-20 keV) spectrum.

\begin{figure*}
  \includegraphics[width=2\columnwidth, angle=0]{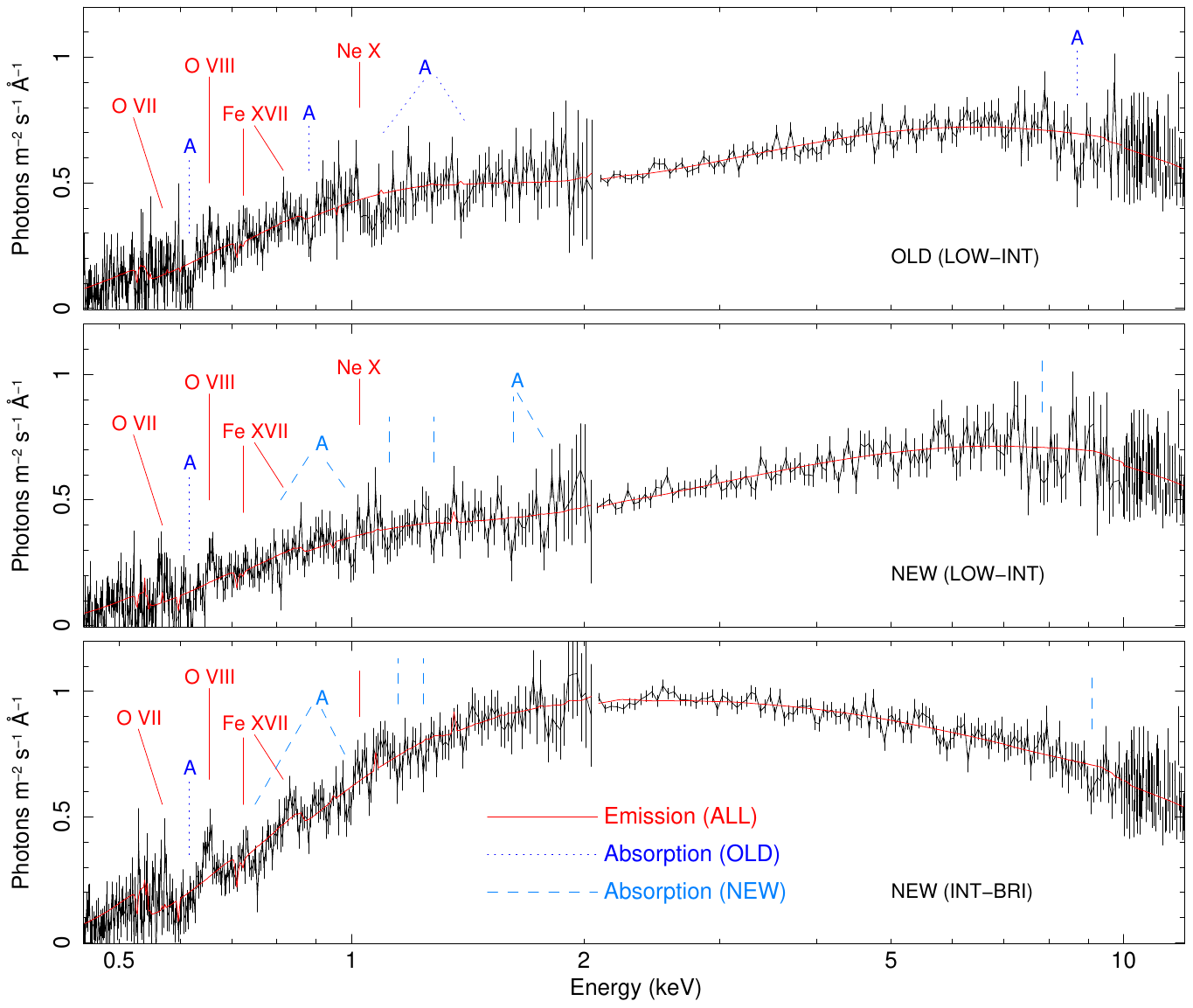}
\vspace{-0.1cm}
   \caption{Flux--resolved XMM-\textit{Newton} RGS (0.45$-$2 keV), EPIC-pn (2$-$10 keV) 
   and \textit{NuSTAR} (10$-$20 keV) spectra.
   The top panel shows the RGS and pn spectra obtained by stacking the archival data 
   (see also \citealt{Pinto2016nature}). The middle panel shows the stacked spectra
   produced using the low-to-intermediate flux observations of our 2017 campaign 
   (ID: 0803990301, 0401, 0701). The bottom panel shows the stacked spectra
   produced using the new intermediate-to-high flux observations 
   (ID: 0803990101, 0201, 0501, 0601); see also Table\,\ref{table:obs_log}.
   The range between 10 and 20 keV is covered by the time-averaged \textit{NuSTAR} spectrum.
   The red lines show the baseline continuum models ($hot(mbb+mbb+comt)$, 
   see Table\,\ref{table:continuum}).
   The strongest emission lines are indicated with solid red labels. 
   Previously known and confirmed absorption features are indicated with dotted blue lines. 
   New possible features are shown as dashed light-blue lines.}
   \label{Fig:Plot_RGS_all}
\vspace{-0.3cm}
\end{figure*}

The flux--resolved RGS and EPIC spectra are shown in Fig.\,\ref{Fig:Plot_RGS_all}.
The top panel shows the archival XMM-\textit{Newton} RGS-EPIC 
spectrum that was also used in \citet{Pinto2016nature}. This spectrum shows 
the canonical low-to-intermediate flux hard ultraluminous state according to the
classification of \citet{Sutton2013}. 
Well known rest-frame emission (and blueshifted absorption) lines are labelled
with solid red (dotted/dashed blue) lines.
The middle panel shows the low-to-intermediate flux spectrum taken during our
new XMM-\textit{Newton} campaign. The spectrum is indeed very similar to the
archival spectrum. 
However, it is clear that some features have \textcolor{black}{weakened (if not disappeared)} 
while some new ones
show up in the new spectrum. Some emission lines, in particular, seem to have
strengthened. 
This suggests long-term temporal variability of the wind and provides another reason 
to avoid the stacking of the old dataset with the new low-intermediate data.
This is even more obvious in the high-state (intermediate-to-bright) 
spectrum of our campaign (bottom panel in Fig.\,\ref{Fig:Plot_RGS_all})
where most emission lines appear stronger and broader than in the low state spectra.

All XMM-\textit{Newton} and \textit{NuSTAR} spectra are sampled 
in channels of at least 1/3 of the spectral resolution, for optimal binning 
and to avoid over-sampling,
and at least 25 counts per bin, \textcolor{black}{using SAS task \textit{specgroup}}. 
However all stacked spectra have 
long exposure times and high quality, which means that such binning affects only the spectral
range at the very low and high energies. We have double checked and found no
effect onto our line or continuum modelling by {decreasing} the binning to simply 1/3
of the spectral resolution.

\begin{table}
\caption{Best-fit broadband continuum models.}  
\label{table:continuum}     
\renewcommand{\arraystretch}{1.3}
 \small\addtolength{\tabcolsep}{-1.pt}
 
\scalebox{1.0}{%
\begin{tabular}{@{} l l l l}     
\hline  
Parameter                                           &  OLD L-I               &  NEW L-I               &  NEW I-B               \\  
\hline                                                                                                                
\multicolumn{4}{c}{\multirow{1}{*}{\textcolor{black}{Baseline Model} : $hot\,(mbb+mbb+comt)$}} \\                                                                                                         
\hline                                                                                        
$L_{X\,mbb1}$                                    &  $3.1 \pm 0.4$      &  $2.4 \pm 0.3$      &  $5.7 \pm 0.4$       \\               
$L_{X\,mbb2}$                                    &  $3.5 \pm 0.5$      &  $3.5 \pm 0.7$      &  $1.5 \pm _{1.5}^{2.5}$       \\  
$L_{X\,comt}$                                     &  $3.7 \pm 0.8$      &  $3.6 \pm 0.9$      &  $7.9 \pm 0.9$     \\  

$kT_{mbb1}$ (keV)                           &  $0.37 \pm 0.02$  &  $0.37 \pm 0.02$  &  $0.5 \pm 0.1$   \\              
$kT_{mbb2}$ (keV)                           &  $1.5  \pm  0.2  $  &  $1.6  \pm  0.2  $  &  $1.3  \pm  0.5  $   \\     
$kT_{in,\,comt}$\,(keV)                        &  $1.5^{(c)}$           &  $1.6^{(c)}$            &  $1.3^{(c)}$            \\                
$kT_{e,\,comt}$\,(keV)                        &  $3.8  \pm  0.2 $   &  $3.9  \pm  0.2 $   &  $4.1  \pm  0.1 $    \\                
$\tau_{comt}$                                    &  $5.1  \pm 0.4 $    &  $5.1  \pm 0.5 $    &  $3.9  \pm 0.1 $   \\                
 
$N_{\rm H}\,(10^{21} {\rm cm}^{-2})$  &   $1.9^{(c)}$           &  $1.9 \pm 0.1$      &     $1.9^{(c)}$        \\  
$C$-stat/d.o.f.                                     &    779/614             &    789/604              &    809/614              \\               
\hline                                                                                                                
\end{tabular}}

$L_{X\,(0.3-20\,\rm keV)}$ luminosities are calculated in $10^{39}$ erg/s, assuming a 
distance of 4.2\,Mpc and are corrected for absorption (or de-absorbed).
The grouped spectra are defined in Sect.\,\ref{sec:fluxed_spectra}.
The best-fit models and the corresponding data are shown in Fig\,\ref{Fig:Plot_RGS_all}.
$^{(c)}$ coupled parameters: for each observation $kT_{in,\,comt}=kT_{mbb2}$, 
while $N_{\rm H}$ is coupled between the observations. The acronyms 
OLD L-I, NEW L-I  and NEW I-B stay for old low-intermediate, new low-intermediate
and new intermediate-bright states. 
\end{table}

\section{Spectral analysis}
\label{sec:spectral_analysis}
 
In this section we present the spectral analysis of NGC 1313 X-1. We first show
the baseline, continuum, model that describes the broadband (0.4$-$20 keV) spectrum.
Then we perform an in-depth analysis of the high-resolution
RGS spectra in order to detect strong sharp features and to search for variability 
in the line strength between different spectral states and epochs.
We then build up self-consistent models of gas in photoionisation 
\textcolor{black}{and collisional-ionisation} equilibrium
to study the ionisation and dynamical structure of the wind responsible for both 
emission and absorption lines.

\subsection{Baseline continuum model}
\label{sec:baseline_continuum}

ULX spectra require several components to obtain
a satisfactory description of the continuum. 
\textcolor{black}{Here we adopt a broadband spectral continuum similar to that one 
used in \citet{Walton2018a}. Two modified blackbody components, \textit{mbb}$_{1,2}$, account
for the outer and inner disc; a comptonised component \textit{comt} describes
the hard tail (see also \citealt{Middleton2015a})}.
All emission components are corrected by absorption due to the foreground interstellar
medium and circumstellar medium using the \textit{hot} model in {\scriptsize{SPEX}} with a low temperature 
$0.5$\,eV (e.g. \citealt{Pinto2013}).
In the spectral fits we couple the column density of the \textit{hot} model 
across all observations
as we do not know physical reasons for which the amount of neutral gas
should significantly change on time scales of a few days.
Moreover, a preliminary spectral fit performed with free column densities 
in the three spectra using the \textit{mbb}$_{1,2}$ + \textit{comt} model
yields consistent column densities.

We apply the $hot(mbb+mbb+comt)$ continuum model 
to the three flux--resolved spectra that we have defined and produced
in Sect.\,\ref{sec:fluxed_spectra}. This phenomenological model provides
a good description of the three broadband spectra. The results 
are shown in Fig.\,\ref{Fig:Plot_RGS_all} and Table\,\ref{table:continuum}.
We obtain an average column density
of $N_{\rm H} = (1.9 \pm 0.1) \times 10^{21} {\rm cm}^{-2}$, which is close to the value
of $(2.5 \pm 0.1) \times 10^{21} {\rm cm}^{-2}$ measured for different continuum 
models in \citet{Walton2020}; see also \citet{Miller2013}.

\begin{figure*}
  \includegraphics[width=2\columnwidth, angle=0]{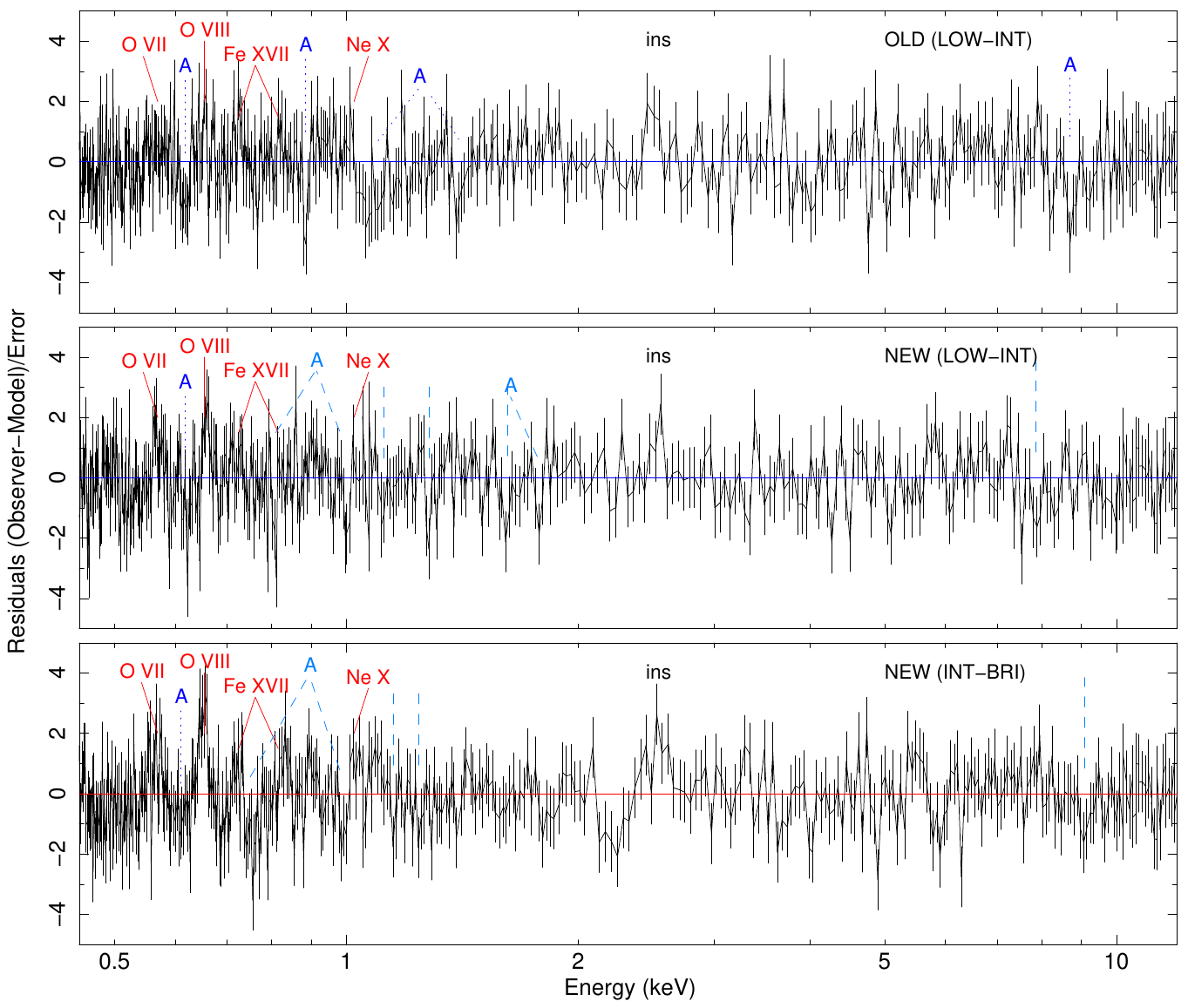}
   \caption{Flux--resolved XMM-\textit{Newton} RGS (0.45$-$2 keV), EPIC-pn (2$-$10 keV) 
   and \textit{NuSTAR} (10$-$20 keV) spectra.
   The plots show the residuals calculated for the spectra in Fig.\,\ref{Fig:Plot_RGS_all} 
   with respect to the baseline continuum models 
   ($hot(mbb+mbb+comt)$, see Table\,\ref{table:continuum}).
   \textcolor{black}{The labels for the lines and the spectra are the same as in Fig.\,\ref{Fig:Plot_RGS_all}.
   The `ins' refers to an EPIC-pn instrumental feature around 2.5 keV.}}
   \label{Fig:Plot_RGS_all_res}
\end{figure*}

The $C$-statistics are high when compared to the 
corresponding degrees of freedom due to strong sharp residuals in the form
of absorption and emission features (see Fig.\,\ref{Fig:Plot_RGS_all_res}) 
as previously shown in \citet{Pinto2016nature}
for the archival data. The best-fit parameters of the models for the old and new 
low-to-intermediate spectra are consistent with each other. 
This means that we can compare these spectra to obtain information on wind long-term
variability at different epochs. The comparison between the bright and low spectra 
of our 2017 campaign instead provide useful constraints on changes
occurring at time scales of about 2-3 months, similarly to the super-orbital 
periods detected in some ULXs (e.g. \citealt{Walton2016b}) and might be associated 
to precession in analogy to that seen in SS433 (e.g. \citealt{Fabrika2004}).


\subsection{Line search with the flux--resolved spectra}
\label{sec:line_search}

Following the approach used in \citet{Pinto2016nature} we search for narrow spectral features
by scanning the spectra with Gaussian lines. We adopt a logarithmic grid of 450 points
with energies between 0.45 (27.5\,{\AA}) and 10 keV (1.24\,{\AA}).
This choice provides a spacing that is comparable to the RGS and pn resolving power 
in the energy range we are investigating 
($R_{\, \rm RGS}\sim100-500$ and $R_{\, \rm pn}\sim20-50$). 
We test the FWHM line widths \textcolor{black}{between} 500 km/s 
(comparable to the RGS spectral resolution)
and 10000 km/s, which is a few times the EPIC resolution. 
At each energy we record the $\Delta C$-stat improvement to the best-fit continuum model
and express the significance (at the energy of the grid) as the square root of the $\Delta C$-stat.
This provides the maximum significance for each line 
\textcolor{black}{(because it neglects the number of trials performed)}. 
We also multiply it by the sign of the gaussian normalisation in order 
to distinguish between emission and absorption features.

\begin{figure*}
  \includegraphics[width=1\columnwidth, angle=0]{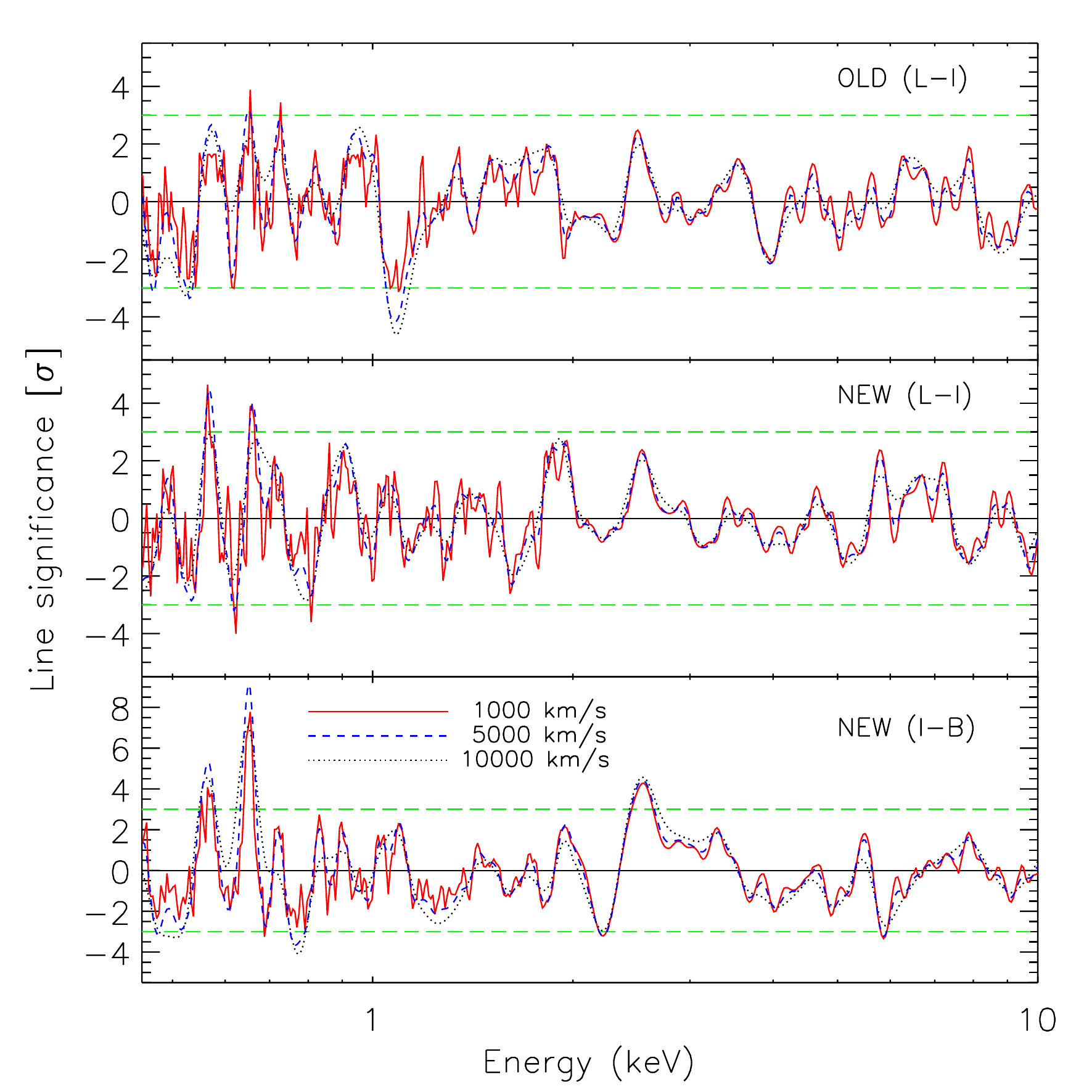}
  \includegraphics[width=1\columnwidth, angle=0]{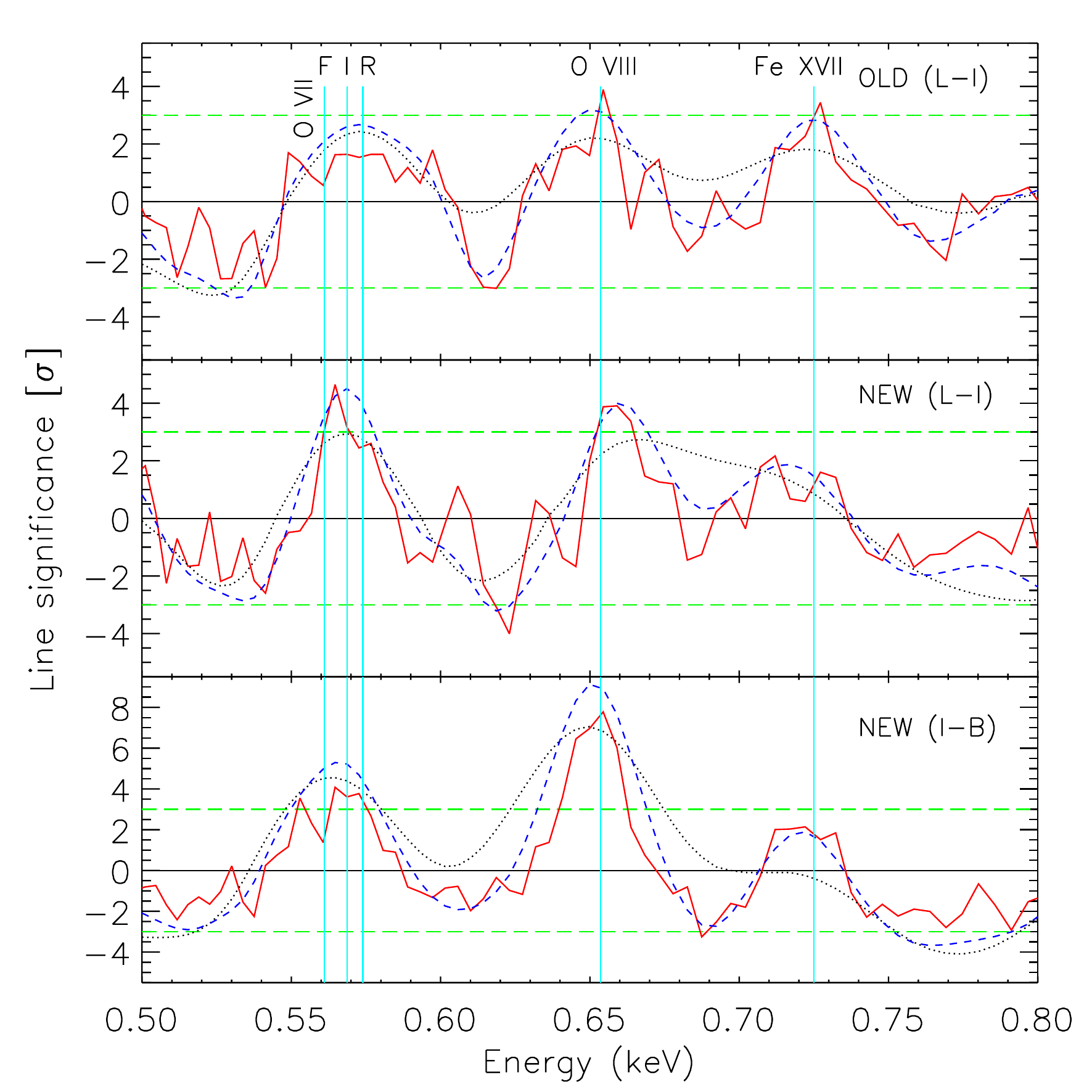}
      \vspace{-0.1cm}
   \caption{Gaussian line scan performed on the three XMM-\textit{Newton} spectra 
   shown in Fig.\,\ref{Fig:Plot_RGS_all} using RGS between 0.45$-$2 keV and EPIC-pn 
   from 2$-$10 keV. The results for three different line widths (FWHM) are shown.
   The line significance is calculated as square root of the $\Delta C$-stat times
   the sign of the gaussian normalisation (positive/negative for emission/absorption lines).
   \textcolor{black}{The right panel is a zoom in the 0.5$-$0.8 keV energy range.}
   The labels `F', `I' and `R' refer to the forbidden, intercombination and resonant lines
   of the {\ovii} triplet.}
   \label{Fig:Plot_line_search}
      \vspace{-0.3cm}
\end{figure*}

In Fig.\,\ref{Fig:Plot_line_search} we show the results of the line scan obtained for the three
RGS+pn spectra using line widths of 1000, 5000 and 10000 km/s. 
The line scan of the archival data once again picks out the strong absorption feature above 
1 keV already reported in \citet{Pinto2016nature} along with other features.
We can notice that such an absorption feature is not significantly detected 
in the new data, but the lower-energy features have strengthened along with the emission lines.
Moreover, the relative strength of the {\ovii} triplet ($\sim$22\,{\AA} or 0.56\,keV) 
and the {\oviii} (19.0 {\AA} or 0.653 keV) emission line seem to have changed. 
This suggests that the ionisation structure of the line-emitting gas 
is variable, which was not clear in previous work, \textcolor{black}{mainly due to the lack of deep RGS observations
from different epochs and spectral states}.


\subsubsection{Variability of the emission lines}
\label{sec:line_search_emission}

\begin{figure}
  \includegraphics[width=1\columnwidth, angle=0]{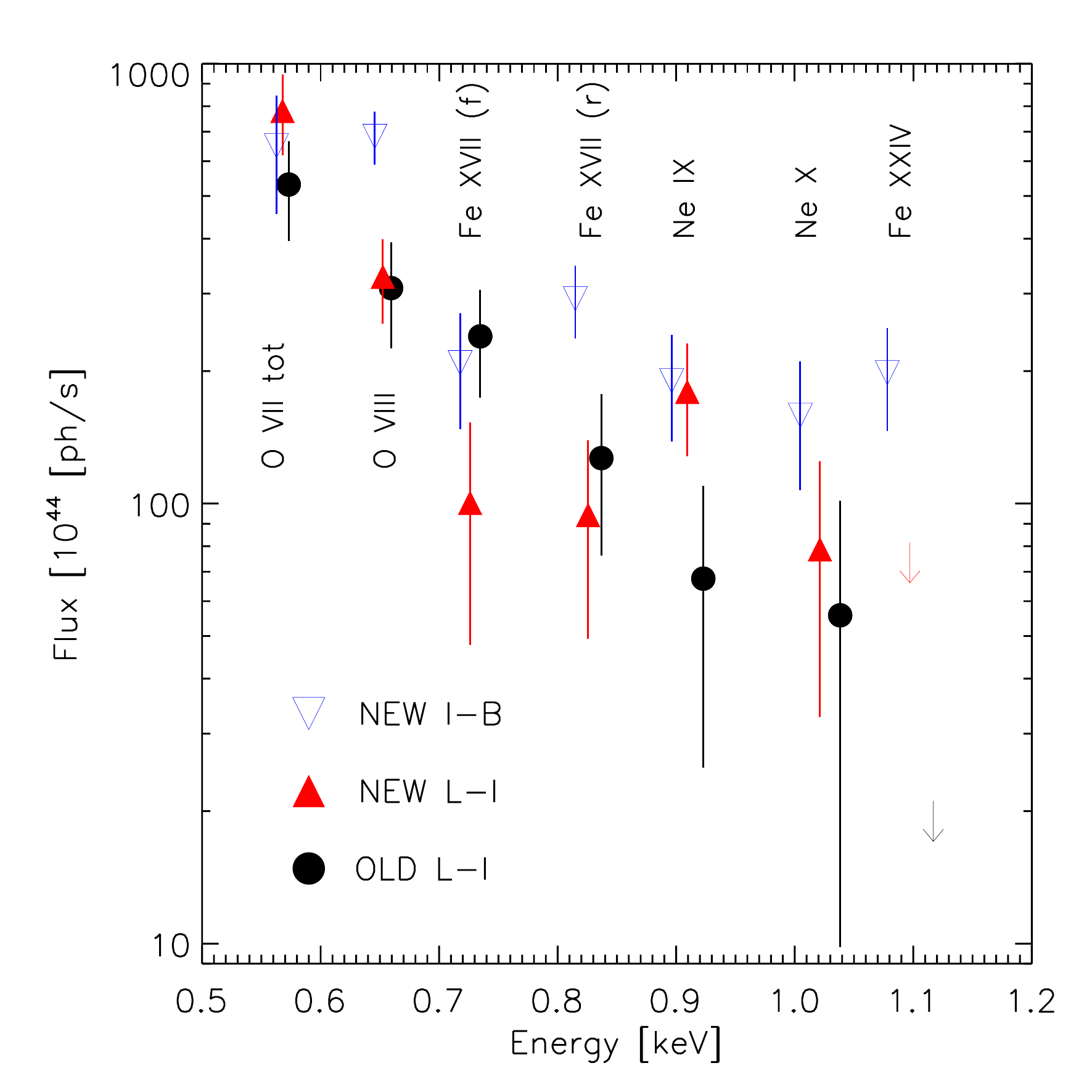}
   \caption{Fluxes of the strongest emission lines in the RGS stacked spectra 
   of NGC 1313 X-1 (new intermediate-bright and low-intermediate 
   and old low-intermediate). The fluxes of the {\ovii} triplet have been accumulated. 
   Notice the significant line variability especially in the new observations.
   The centroids of the lines have been slightly shifted for displaying purposes.
   {In two cases only 90\% upper limits could be obtained.}}
      \label{Fig:Plot_Gaussians}
      \vspace{-0.3cm}
\end{figure}

We compare the observed fluxes of the strongest emission lines detected with the Gaussian scan for
the three flux-resolved spectra in Fig.\,\ref{Fig:Plot_Gaussians}. 
Here, we use the results of the broad
core (FWHM=5000 km/s), which maximises the {\ovii} and {\oviii} significance.
Owing to low statistics above 20\,{\AA} and the large line width, 
the {\ovii} triplet is not well resolved.
We therefore prefer to accumulate the fluxes of the {\ovii} lines (resonance $21.6$\,{\AA},
intercombination $21.8$\,{\AA} and forbidden $22.1$\,{\AA} lines).

It is clear that the emission lines significantly change, particularly during the 
bright state, but it is not straightforward to understand the variability of the emission measure 
distribution by simply comparing the single gaussian lines (Fig.\,\ref{Fig:Plot_Gaussians}).
However, the largest changes are seen in the {\oviii} and other high-ionisation lines,
whose brightening might suggest a higher ionisation state in the \textcolor{black}{brighter spectrum}.
A self-consistent model is required to tie the results from multiple lines together and obtain 
further insights into the gas dynamics. This is done in Sect.\,\ref{sec:emitting_gas}
where we build up photoionisation and collisional-ionisation emission models
and perform model-scans onto the data.

\begin{figure}
  \includegraphics[width=1\columnwidth, angle=0]{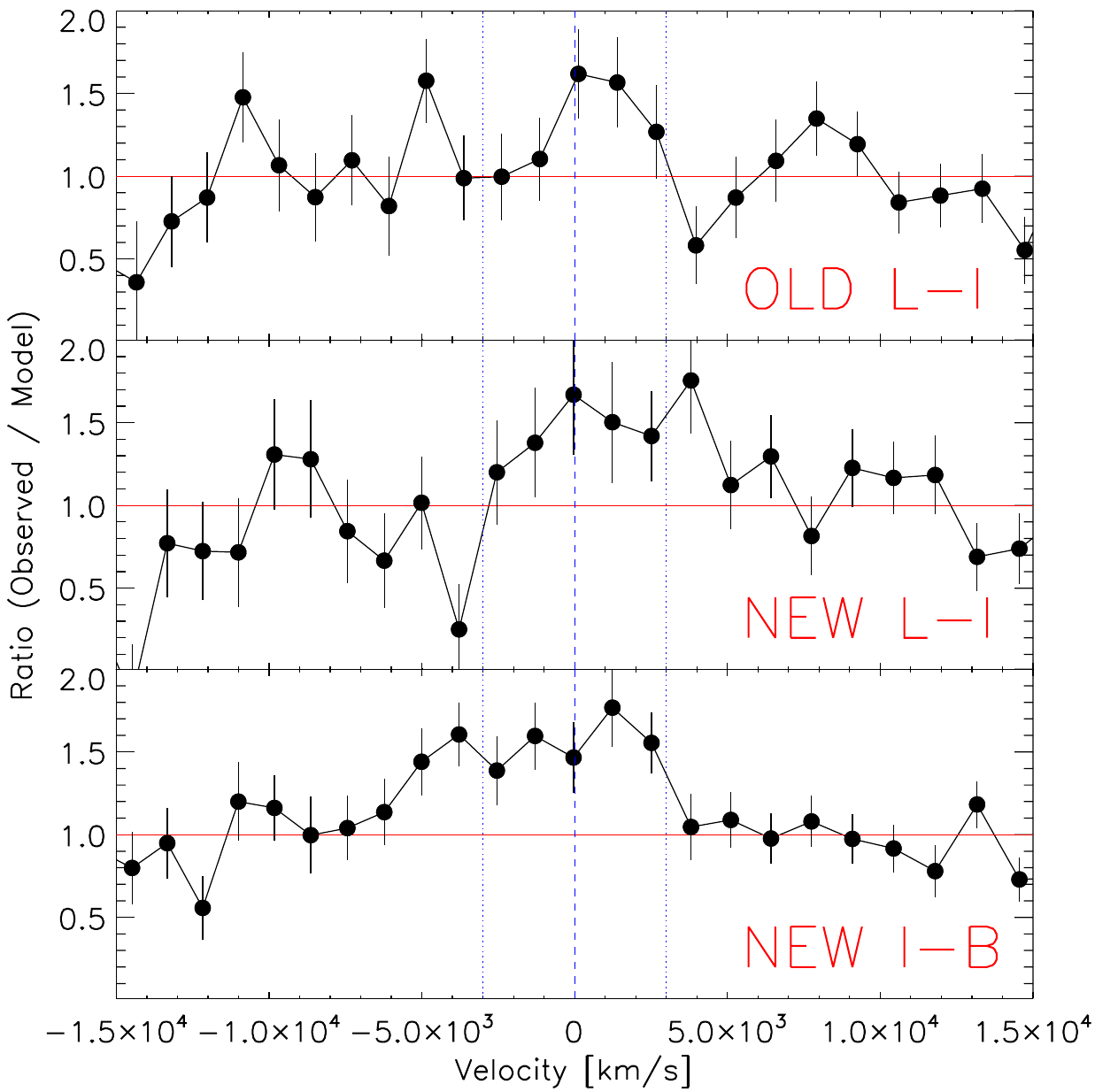}
   \caption{{\oviii} Lyman $\alpha$ ($1s-2p$) velocity profile for the three RGS flux-resolved 
   spectra of NGC 1313 X-1 (see also Fig.\,\ref{Fig:Plot_RGS_all}).
   A positive velocity indicates gas motion towards the observer.
   A line centroid of 18.967\,{\AA} ($\sim0.654$\,keV) is adopted (blue dashed line).
   The dotted vertical lines indicate shifts of $\pm\,3000$ km/s.
   In the brighter state the line seems to exhibit a stronger red wing.}
      \label{Fig:Plot_Gaussians_OVIIIvel}
      \vspace{-0.3cm}
\end{figure}

It is difficult to determine whether the emitting gas is in collisional or photoionisation 
equilibrium owing to the limited statistics in each line, but some diagnostics can be
obtained using the following well-known ratios:
\begin{equation}\label{eq:R_ratio}
r = \frac{F}{I}
\end{equation}
\begin{equation}\label{eq:G_ratio}
g = \frac{F+I}{R}
\end{equation}
where $F$, $I$ and $R$ are the fluxes of the forbidden, intercombination and resonance
lines. By combining the results from all three spectra we can place some constraints 
which would be difficult, if not impossible, to do for each individual spectrum.
Accounting for uncertainties in the 
line widths, the continuum and the RGS calibration, we measure
for the {\ovii} triplet the following ratios: 
$r = 0.6\pm_{0.4}^{1.0}$ and $g=1.3\pm_{0.8}^{1.6}$. The uncertainties on these 
ratios are still large, but their values suggest
a high-density hybrid gas ($n_{\rm H} \sim 10^{10-12}$ cm$^{-3}$) where both
recombination and collisional processes occur \citep{Porquet2000}.

\vspace{0.6cm}

The emission lines are broadly consistent with being at their rest wavelengths, although the 
\textcolor{black}{centroids} of the {\ovii} and {\oviii} lines in the bright state seem to be redshifted by
$\sim 3000$ km/s with respect to those of the low state
as clearly shown in Fig.\,\ref{Fig:Plot_Gaussians_OVIIIvel} for the {\oviii} Lyman\,$\alpha$ line.

\subsection{Photoionisation modelling}
\label{sec:photoionisation_modelling}

An accurate photoionisation model requires the knowledge of the radiation field.
\citet{Pinto2019a} have studied the thermal stability of ULX winds and compared them
with other high Eddington sources such as Narrow Line Seyfert 1 galaxies.
In that paper the spectral energy distribution (SED) of NGC 1313 X-1 was computed
from IR to hard X-ray energies using literature data for the IR/optical/UV domain,
complemented with the optical monitor aboard XMM-\textit{Newton}.
The X-ray spectral portion is provided by XMM-\textit{Newton} and \textit{NuSTAR}.

\subsubsection{SED and stability curves}
\label{sec:SED}

\begin{figure}
  \includegraphics[width=1\columnwidth, angle=0]{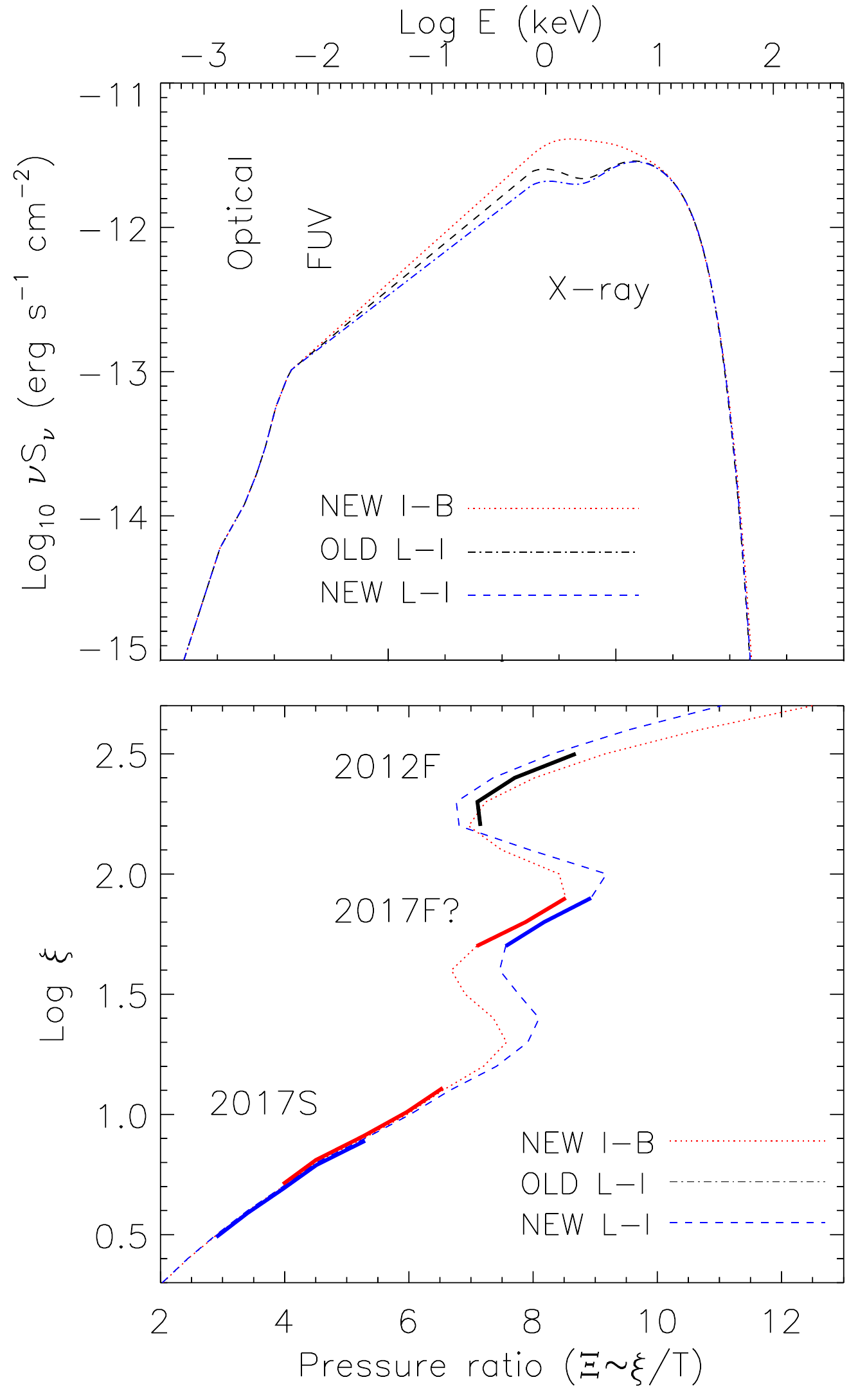}
  \vspace{-0.6cm}
   \caption{Spectral energy distribution (top) and thermal stability curve 
   (ionisation parameter versus radiation/thermal pressure ratio, bottom) 
   computed for the three spectra under investigation (see also \citealt{Pinto2019a}). 
   The solid lines in the bottom panel indicate the ionisation 
   parameters ($\pm1\sigma$ errors) 
   of the photoionised absorbers with higher significance in the archival (2012 \textcolor{black}{at latest})
   and in the new (2017) spectra \textcolor{black}{(see Table\,\ref{table:improvements})}. 
   The labels ``F'' and ``S'' refers to the fast and slow components,
   while ``F?'' indicates the component with lower significance ($\lesssim3\sigma$)
   in the new data.}
   \label{Fig:Plot_SED_Balance}
  \vspace{-0.4cm}
\end{figure}

In this work we construct the SEDs of the low-to-intermediate and intermediate-to-bright states
using the low-energy portion (IR, optical and far-UV) of the SED computed by \citet{Pinto2019a}
for the time-averaged spectrum of NGC 1313 X-1.
For the high-energy part (soft and hard X-rays from 0.4 to 20 keV) we use the best-fit continuum 
model, $hot(mbb+mbb+comt)$, estimated in Sect.\,\ref{sec:baseline_continuum}, 
Fig.\,\ref{Fig:Plot_RGS_all} and Table\,\ref{Fig:Plot_RGS_all}.
The absorption from the ISM is removed to obtain the intrinsic source spectrum.
As shown in the spectral modelling, the SED above 10 keV is taken to be constant
and is provided by the time-averaged \textit{NuSTAR} spectrum.
In fact, \citet{Walton2020} show that there is very little variability above 10 keV
in the \textit{NuSTAR} spectra of NGC 1313 X-1.
The final SEDs are shown in Fig.\,\ref{Fig:Plot_SED_Balance} (top).

The study of the relationship between the UV and X-ray flux will be done elsewhere,
as well as a detailed time-resolved study of NGC 1313 X-1 comparing individual exposures
or nearby pairs of exposures. For this work we adopt a constant flux at the low energies
(IR, optical and far-UV). A study of some systematic effects introduced by the variability
of the low-energy portion of the spectrum onto the ionisation balance is briefly discussed
in \citet{Pinto2019a}. They show that the main effect of a lower IR-to-UV flux is that of
strengthening thermal instabilities in the plasma at intermediate temperatures and 
ionisation parameters (e.g. Log T ({\rm K}) $\sim 5$ and Log $\xi \, ({\rm erg/s \, cm}) \sim 1.5$,
respectively), but are not expected to produce major effects to our results.

Plasmas in photoionisation equilibrium are characterized by the ionisation parameter, $\xi$, 
defined as (see, e.g., \citealt{Tarter1969}):

\vspace{-0.1cm}         
\begin{equation} \label{Eq:Eq_ionpar}
\xi = \frac{L_{\rm ion}}{n_{\rm H} \, R^2}
\end{equation}
where $L_{ion}$ is the ionising luminosity (taken between 1 and 1000 Rydberg,
i.e. 13.6 eV and 13.6 keV), $n_{\rm H}$ the hydrogen number density and 
$R$ the distance from the ionising source. 
The ionisation balance slightly depends on the abundances; here we adopt
the recommended proto-Solar abundances of \citet{Lodders2009}, 
which are default in {\scriptsize{SPEX}}.
Although ULXs are often found in star-forming galaxies where non-Solar metallicity
might be expected, currently there does not seem to be strong 
evidence for highly non-solar abundances for the ULXs themselves 
($Z\sim0.6-0.7 Z_{\odot}$, 
see, e.g., \citealt{Winter2007}, \citealt{Mizuno2007} and \citealt{Pintore2012}). 
Moreover, the evidence of sub-Solar abundances in the ULX 
host galaxy (or along the line of sight) does not necessarily indicate
the abundances in the accretion disc \citep{Prestwich2013}.

We calculate the ionisation balance using {\scriptsize{SPEX}}
newly implemented \textit{pion} code, which is optimised to perform 
instantaneous calculation of ionisation balance in the regime of photoionisation.
We also test the \textit{xabsinput} tool in {\scriptsize{SPEX}}, which performs
the same calculation and provides in output the ionisation balance in a format
useful for several photoionisation absorption models (see Sect. \ref{sec:absorbing_gas}).
The results obtained with \textit{pion} and \textit{xabsinput} are consistent.
Following the method used in \citet{Pinto2019a} for the stability (or $S$) curve
of the low state SED, we compute the ionisation balance for the high state. 
The $S$ curves show the ionisation parameter, $\xi$,
as a function of the ratio between the radiation pressure ($F/c$) and thermal pressure 
($n_{\rm H}kT$), given by 

\vspace{-0.1cm}         
\begin{equation} \label{Eq:Eq_pressure}
\Xi = \frac{F}{n_{\rm H} c kT} = 19222 \, \frac{\xi}{T},
\end{equation}
with $F = L / 4 \pi r^2$ \citep{Krolik1981}.
The stability curves computed for the three SEDs are shown
in Fig.\,\ref{Fig:Plot_SED_Balance} (bottom panel). 
Along these curves 
heating equals cooling and, therefore, the gas is in thermal balance.
Where the $S$ curve has a positive gradient, 
the photoionised gas is thermally stable or, in other words, small perturbations 
upwards (downwards) will be balanced by an increase of cooling (heating).

The stability curves of the three spectra under investigation are very similar,
which means that thermal instabilities should not affect the evolution of the wind.
The study of thermal stability in each observation is relevant but
is beyond the scope of this paper and will be done elsewhere.

{We have tested the effect of non-Solar metallicity by calculating the ionisation
balance for $Z$ between 0.5 and 1.5. 
Mildly sub-Solar ($\sim$ 0.7) metallicities have very small effects onto 
the ionisation balance with shifts comparable to those observed in Fig.\,\ref{Fig:Plot_SED_Balance}.
In particular, sub-Solar metallicities tend to increase the stable brunch.
For $Z \gtrsim 1.5$ the unstable portion of the $S$ curves increases.
We therefore do not expect significant issues in our ionisation balance
calculation and spectral fits due to the uncertainties in the abundance pattern 
(see also \citealt{Pinto2012b}).}

\begin{figure}
 \centering
  \includegraphics[width=0.99\columnwidth, angle=0]{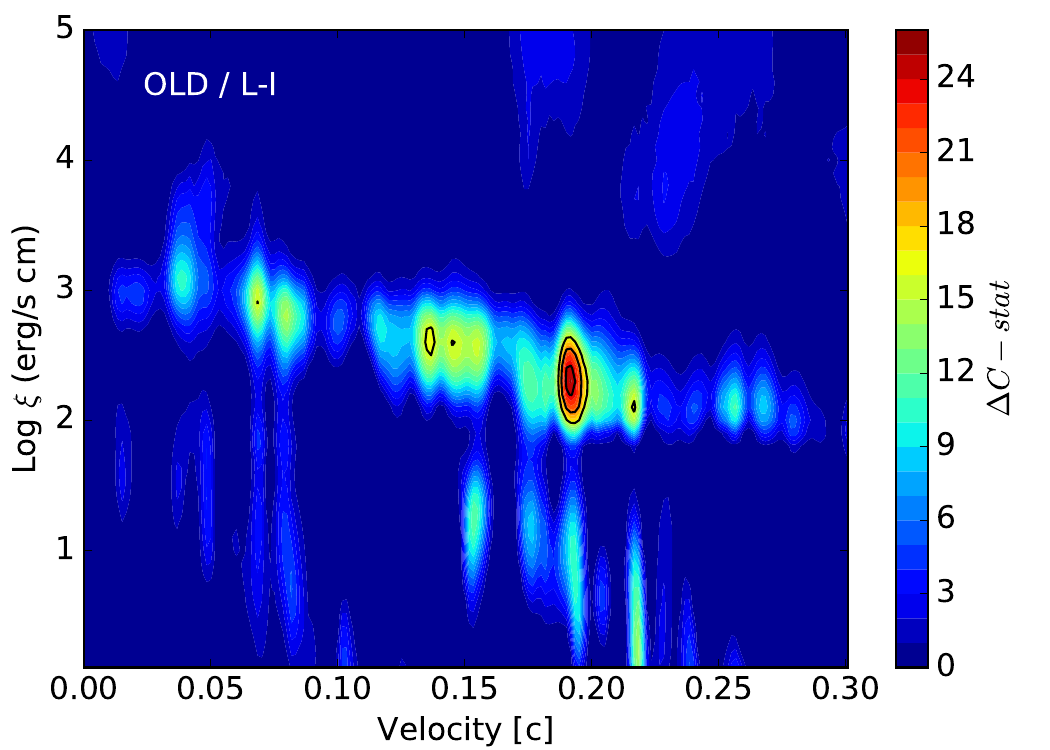} 
  \includegraphics[width=0.99\columnwidth, angle=0]{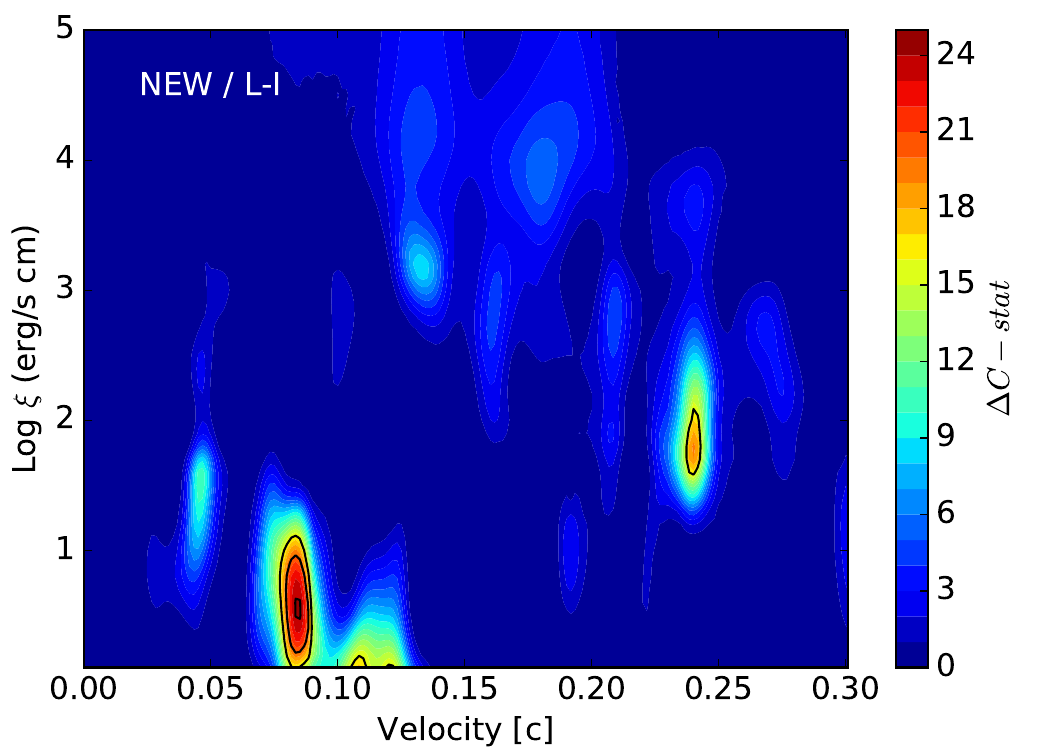}
  \includegraphics[width=0.99\columnwidth, angle=0]{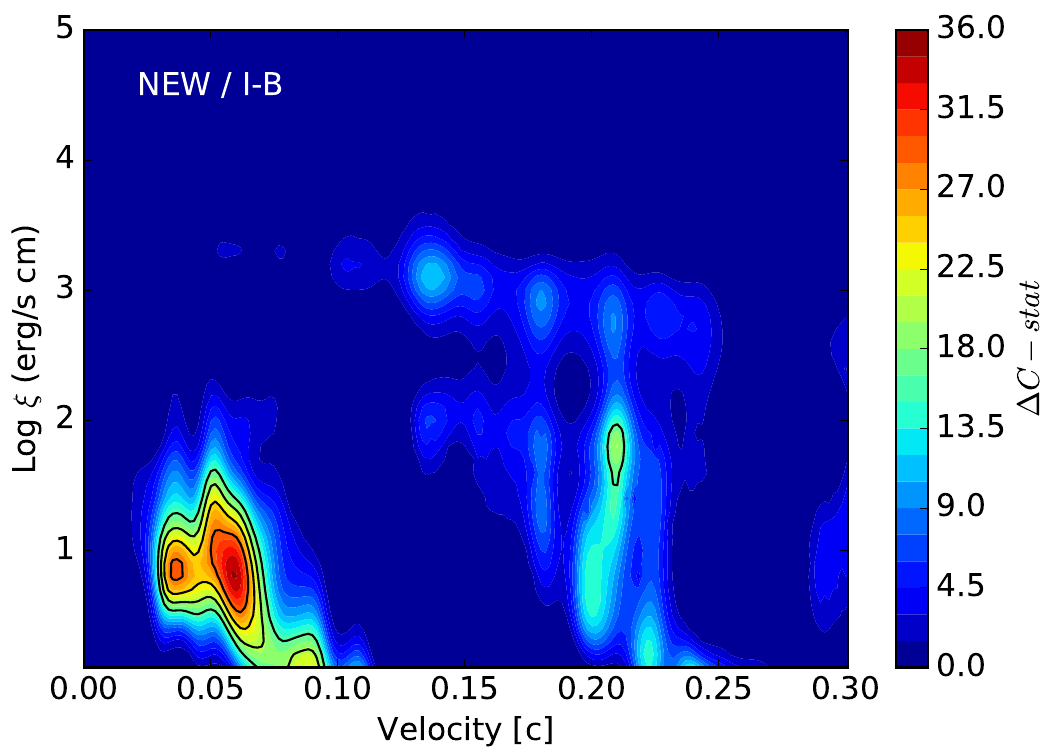}
   \caption{\textcolor{black}{Photoionisation absorption model scan of the archival data (top),
                the new low-to-intermediate (middle) and the new intermediate-to-bright
                spectra (bottom). 
                The color is coded according to the $\Delta C$-stat fit improvement 
                to the continuum model. The black contours refer to the 2.5, 3.0, 3.5
                and 4.0\,$\sigma$ confidence levels estimated through 3\,x\,10\,000
                Monte Carlo simulations.}}
   \label{Fig:absorbing_gas}
  \vspace{-0.4cm}
\end{figure}


\subsubsection{Absorbing gas}
\label{sec:absorbing_gas}

Physical models are useful to determine the significance and nature of 
spectral features and to study plasma dynamics.
We first focus on the modelling of the absorption features by using the \textcolor{black}{fast} $xabs$ model
in {\scriptsize{SPEX}}. $xabs$ calculates the transmission of a slab of material, 
where all ionic column densities are linked through a photoionisation model. 

In principle, one could fit for several parameters of the $xabs$ model such as
the column density, $N_{\rm H}$, the ionisation parameter, $\xi$,
the line-of-sight velocity, $v_{\rm LOS}$, the velocity dispersion,
the abundances, etc. However, in order to determine the best solution based 
on the observational dataset, we perform automated spectral scans with the 
$xabs$ model using a pre-defined multidimensional grid, 
similar to \citet{Kosec2018b}. This technique is very efficient at locating 
possible solutions and prevents the fits from getting stuck in local minima, 
although is computationally expensive. 
It also provides a proxy for the absorption (emission) measure distribution 
if applied to line absorption (emission) models.

In this case, we construct a multidimensional grid of $v_{\rm LOS}$ and
$\xi$, whilst keeping free the $N_{\rm H}$. 
We adopted a $v_{\rm LOS}$ blueshift  
ranging from $0$ to $0.3c$ with steps of 500 km/s
(comparable to the spectral resolution of the highest-resolution instrument, RGS).
A grid of $\log \xi$ from 0.1 to 5.0 with a 0.1 step is also used.
We also do not fit the line widths but rather assume a grid of specific values 
similarly to the line scan in Sect.\,\ref{sec:line_search}
($\sigma_{v,1D} = 1/2.355$ FWHM of 500, 1000, 5000 km/s).
All element abundances are chosen to be Solar
for the same reasons mentioned in Sect. \ref{sec:SED}.
Freeing the abundances could possibly provide significant
improvements to the model search, but also increase the number of trials
and the computing time.
For every $(\xi, v_{\rm LOS}, \sigma_v)$ combination a spectral fit is performed
to each of the three flux-resolved spectra,
starting from the baseline continuum model defined in Sect.\,\ref{sec:baseline_continuum},
with the continuum parameters and the column density of the absorber free to vary,
and the statistical fit improvement in $\Delta\,C$ is recovered, 
as done for the Gaussian line scan in Sect.\,\ref{sec:line_search}. 

In Fig.\,\ref{Fig:absorbing_gas} we show the \textcolor{black}{probability distributions 
(in $\Delta\,C$ and $\sigma$) from scans of the three spectra with $\sigma_v = 500$ km/s
for the archival data and 1000 km/s for the two new spectra,
which yield strong detections of both emission and absorption lines.}
As expected, our code finds and confirms the best-fit $\sim0.2c$ solution 
previously found in \citet{Pinto2016nature}. 
Additional, weaker, peaks are found at lower ionisation states 
($\log \xi \sim 0.6-0.8$) and velocities ($0.06-0.08c$).

As anticipated by the line scan in Fig.\,\ref{Fig:Plot_line_search}, 
the model scan shows that the absorber has radically changed, with the bulk of the absorption
occurring at lower velocities (see Fig.\,\ref{Fig:absorbing_gas}) in the new data.
These velocities agree with those suggested by 
principal component analysis (see Sect.\,\ref{sec:timing_analysis}).
Secondary peaks ranging from $0.2-0.25c$ with higher $\xi$ similar to the archival 
data are also found in the new data albeit at much lower significance.

\textcolor{black}{The two components with low or high ($\xi$, $v$) 
are likely independent as they account for different 
features in the RGS spectra. Although their individual significance is lower
when both components (and the model for the line emission)
are present in the model, they still provide substantial improvements to the
fits (see Appendix\,\ref{sec:appendix}, Table\,\ref{table:improvements} and Fig.\,\ref{Fig:rgs_spectrum_fit}).}

We note that the global $\Delta\,C$ levels are lower than those quoted 
in \citet{Pinto2016nature} because here we do not use the EPIC data below 2 keV,
which has a larger count rate but lower spectral resolution than RGS, 
by more than an order of magnitude, in order to decrease the degeneracy.
We have done a quick check by temporarily including pn data between 0.3-2 keV and
obtained $\Delta\,C$ of up to 48 with respect to the continuum (no absorption model)
for the archival (OLD L-I) data and $\Delta\,C$ up to 36 and 83 for the new spectra
(NEW L-I and NEW I-B, respectively). \textcolor{black}{These values correspond to the highest peaks
shown in Fig.\,\ref{Fig:absorbing_gas}}.

{As can be seen in Fig.\,\ref{Fig:absorbing_gas}, our data selection is not optimal for plasmas 
with log $\xi \gtrsim 4.0$. The significance and thorough modelling of the putative, weaker, absorption 
features in the Fe K energy band (see Fig.\,\ref{Fig:Plot_RGS_all_res}) will require to combine
data from EPIC and FPM-A/B as previously done in \cite{Walton2016a} and
will be focus of a forthcoming paper.}

\subsubsection{Emitting gas}
\label{sec:emitting_gas}

\begin{figure}
  \includegraphics[width=0.99\columnwidth, angle=0]{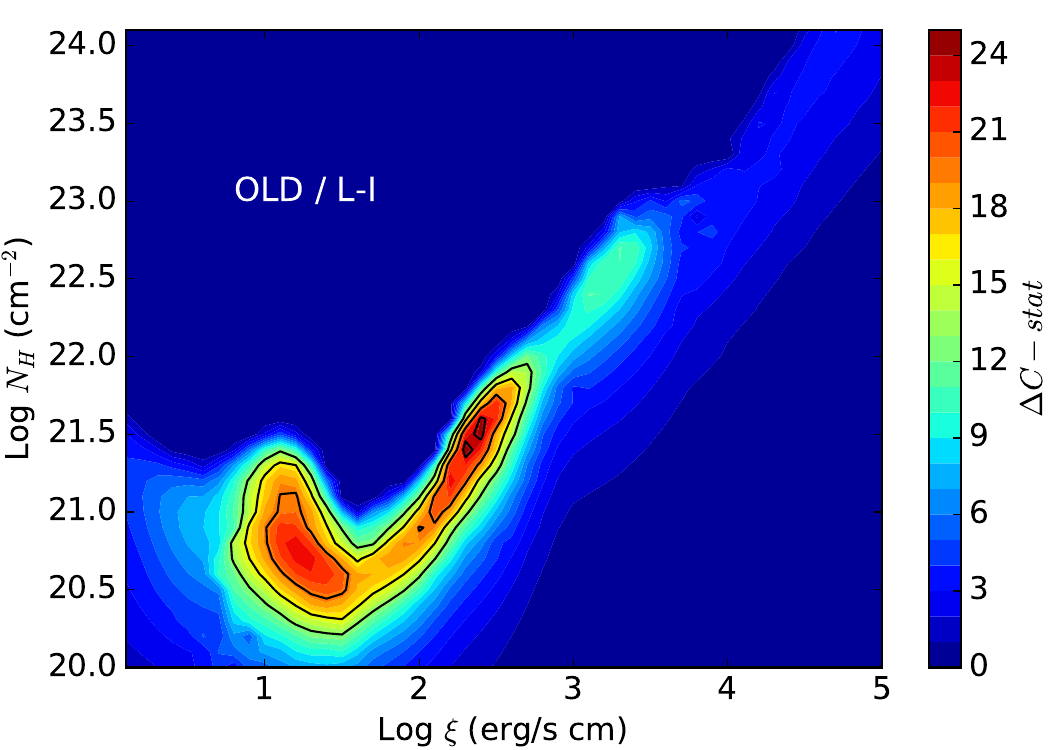} 
  \includegraphics[width=0.99\columnwidth, angle=0]{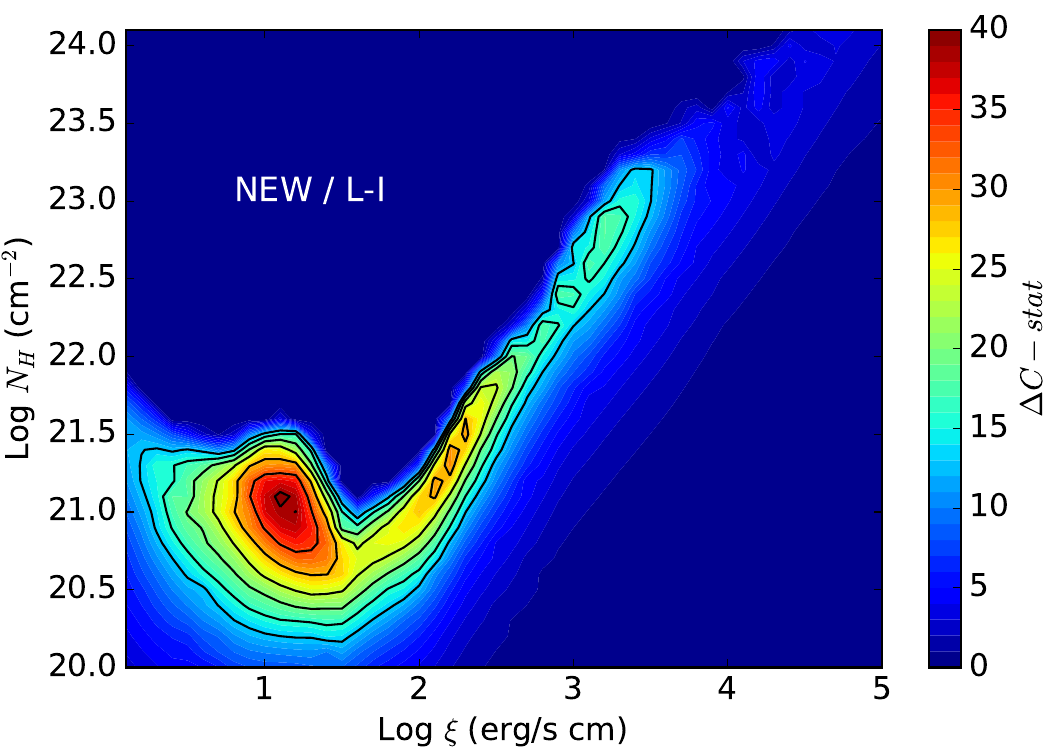}
  \includegraphics[width=0.99\columnwidth, angle=0]{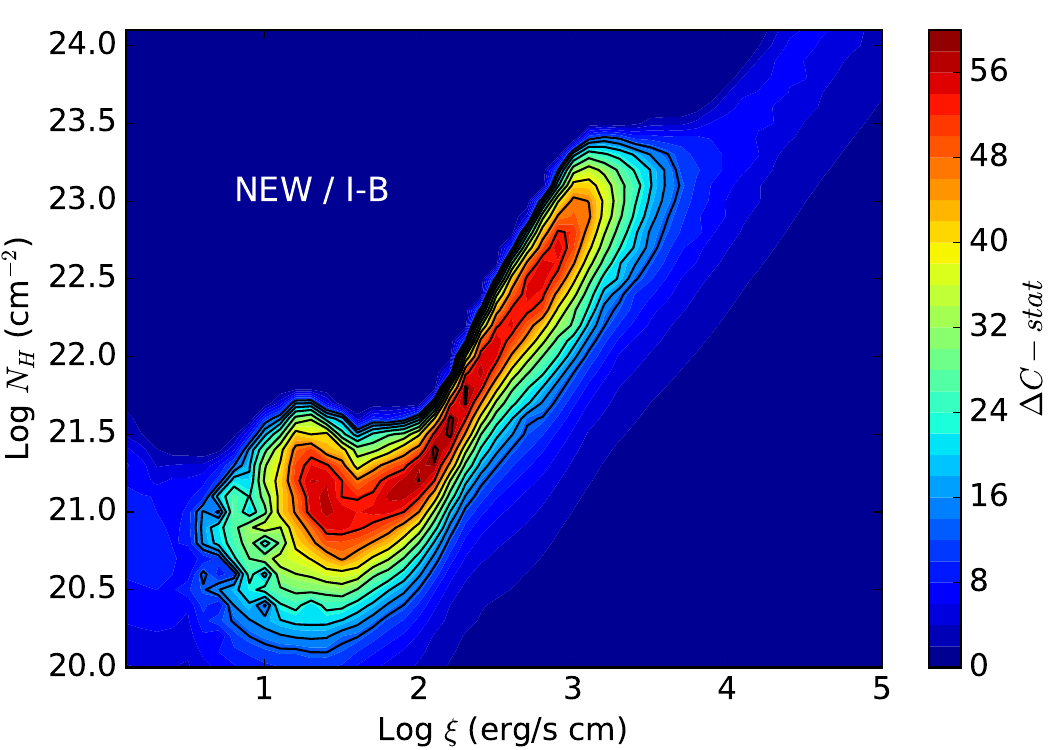}
   \vspace{-0.5cm}
  \caption{\textcolor{black}{Photoionisation emission model scan of the archival data (top),
   the new low-to-intermediate spectra (middle) and the new intermediate-to-bright
   spectra (bottom panel). 
                The color is coded according to the $\Delta C$-stat fit improvement 
                to the continuum model. The black contours refer to the 2.5, 3.0, 3.5,
                4.0\,$\sigma$,  etc. confidence levels estimated through 10\,000
                Monte Carlo simulations for each spectrum. Negative values of $\Delta C$-stat for
                high $N_{\rm H}$ at low-to-mild $\xi$ are kept to zero for 
                displaying purposes.}}
   \label{Fig:emitting_gas}
   \vspace{-0.4cm}
\end{figure}


In order to determine the emission measure distribution of the plasma in the three 
spectra, we perform a similar model scan through the \textit{pion} model in {\scriptsize{SPEX}}.
The \textit{pion} model calculates the transmission and emission of a photoionised plasma,
whilst the \textit{xabs} model is only optimized for absorption. The photoionisation
equilibrium in this model is calculated self-consistently. In principle, we could just use
\textit{pion} for both emission and absorption. However, this model re-calculates the 
ionisation balance at every iteration and therefore is much more computationally 
expensive than the \textit{xabs} model. For this reason we prefer to use it only
for emission studies. The \textit{pion} model shares several parameters with \textit{xabs},
but it also allows us to fit the density \textcolor{black}{(for line emission)}
and retrieve more information on the plasma properties.

The emission lines detected with the Gaussian line scan are consistent with being
at their rest-frame wavelengths (Fig.\,\ref{Fig:Plot_Gaussians}), which agrees with
what was previously found in \citet{Pinto2016nature}. We therefore build a simple grid
of \textit{pion} models with $\xi$ between 0.1 and 5.0 with 0.1 steps
\textcolor{black}{and $N_{\rm H}$ between $10^{20-24}$ cm$^{-2}$}.
\textcolor{black}{Based on the results of the line scan and the {\ovii} triplet shape discussed in 
Sect.\,\ref{sec:line_search_emission}}, 
we adopt the plasma density to be equal to $n_{\rm H} = 10^{10}$ cm$^{-3}$ and 
the line width $\sigma_v = 1000$ km/s. 
We prefer not to increase the $n_{\rm H}$ further as there are some issues 
with certain level populations at very high densities in \textit{pion}.
These $\xi$ grids are used for the \textit{pion} model, which is then an additive
component in our spectral model on top of the continuum model
(the \textit{xabs} absorption component is removed from this model).
For the \textit{pion} model we also adopt covering fraction equal to zero 
\textcolor{black}{(i.e using \textit{pion} only to produce emission lines)} 
and solid angle $\Omega=4\pi$.
Fitting additional parameters such as $\Omega$ might provide even better fits but 
would significantly increase the computing time.
\textcolor{black}{Zero line-of-sight velocity is adopted here.}

We run the model scan for the three spectra in investigation and plot the results
in Fig.\,\ref{Fig:emitting_gas}. \textcolor{black}{Each panel shows the probability distribution
of the emitting gas in terms of $\Delta C$-stat and the confidence level expressed 
in $\sigma$, which is constrained using Monte Carlo simulations
(i.e. including look-elsewhere effects). 
As we predicted above, the main peaks
of the distributions show up at lower ionisation parameters during the 2017 observations,
particularly in the low flux spectrum. However, the picture seems rather complex
with evidence for multiphase structure and larger fractions of hotter gas during the high state.}

\section{Principal component analysis}
\label{sec:timing_analysis}

In this section we use an alternative technique to search for time-dependent spectral features 
at lower spectral resolution using principal component analysis (PCA) and the high-statistics EPIC-pn
spectra to search for spectral features that respond to continuum variations.


In the field of active galactic nuclei, principal component analysis (PCA) has proven
to be a reliable tool to detect broadband or narrow spectral features that have
correlated variability (see, e.g., \citealt{Mittaz1990}, \citealt{Francis1999}, \citealt{Vaughan2004}, \citet{Miller2007pca}, \citealt{Miller2008pca}, \citealt{Parker2014}).
For instance, \citet{Parker2017b} used PCA to detect the spectral component
of the ultrafast ($\sim0.2c$) outflow in Narrow Line Seyfert 1 IRAS 13224$-$3809
thanks to the fact that the wind responds to the variations in the source continuum  
(\citealt{Parker2017a}, \citealt{Pinto2018a}).
They identified a series of variability peaks in both the first PCA component
and $F_{\rm var}$ spectrum, which correspond to the strongest predicted absorption lines
from the ultrafast outflow previously found in high-resolution X-ray spectra. 
The main requirements for this method to be successful are 
a) availability of a large dataset, b) source variability at time scales larger 
than the time slices used (typically 1-10 ks) and 
c) wind and continuum correlated variability.

\begin{figure}
  \includegraphics[width=1\columnwidth, angle=0]{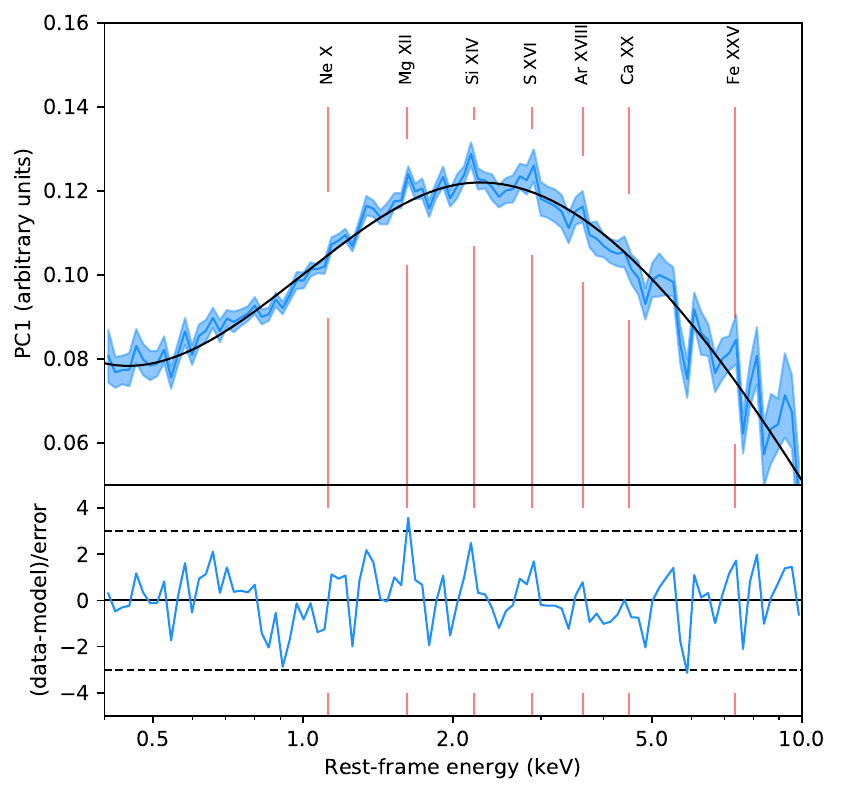}
  \vspace{-0.3cm} 
   \caption{XMM/EPIC-pn PCA 
   \textcolor{black}{1st order spectrum, calculated from the entire dataset (top). 
   The main relevant transitions, blueshifted by 0.08$c$, are marked by vertical lines, and 
   the shaded area shows the 1$\sigma$ errors. 
   The black line shows a spline fit to these data. 
   The bottom panel shows the residuals to this fit, in units of standard deviations. 
   Dashed horizontal lines mark the 3$\sigma$ levels.}}
   \label{Fig:Plot_PCA}
  \vspace{-0.3cm} 
\end{figure}

We apply this technique, using the code of \citet{Parker2017a}, 
to the whole NGC 1313 X-1 EPIC-pn data in order to search for 
similar peaks (all 12 observations in Table\,\ref{table:obs_log}). 
We adopt 10 ks time bins and a logarithmic energy grid
for a total of 165 pn spectra. The results for principal component one, PC1, 
are shown in Fig\,\ref{Fig:Plot_PCA}.
Some features appear at intermediate energies (1$-$3 keV) that 
would correspond to a $0.08c$ outflow if we identify them with the strongest transitions
in this energy band (also detected in IRAS 13224$-$3809). 
At high energies noise dominates and it is difficult to obtain further information, 
while at low energies either PCA is not very sensitive to the features or they are
weakly correlated to the continuum or the variability is weaker in the soft band.
There is an interesting tentative drop at 5.9 keV similar to the bright spectrum 
(Fig.\,\ref{Fig:Plot_line_search}).

\textcolor{black}{We fit the PCA spectrum with a cubic spline, using the \textsc{UnivariateSpline} function 
from \textsc{scipy}, weighting the points by the inverse of their errors. This gives a reasonable 
description of the broadband shape, which we can then use to estimate the line significances. 
For the transitions that have clear corresponding peaks in the PCA spectrum (Ne\,{\sc{x}}, 
Mg\,{\sc{xii}}, Si\,{\sc{xiv}}, S\,{\sc{xvi}} and Ar\,{\sc{xviii}}), we calculate their deviation from the spline in 
$\sigma$ (1.2, 3.6, 2.5, 1.7, and 0.8, respectively). We convert these to probabilities, assuming Gaussian
statistics, and multiply them together to calculate the probability of finding five lines of this strength at the 
same blueshift. We then account for the number of trials by dividing by the number of resolution elements 
over the relevant range of velocities (for $v=0$ to $v=0.4c$, at 2~keV, this gives $\sim8$ resolution 
elements). This gives a final chance probability of $2.75\times10^{-7}$, or 5.0$\sigma$. Considering only 
the strongest and most obvious three features (Mg\,{\sc{xii}}, Si\,{\sc{xiv}}, S\,{\sc{xvi}}) this falls to 
$P=2.8\times10^{-6}$, or 4.5$\sigma$.}

We attempt to repeat the PCA analysis using only the new data from the 2017
campaign and obtained consistent results albeit with higher noise.
The archival data alone, where strong absorption was detected above 1 keV,
is not ideally deep for our PCA analysis, but a quick check shows a weak 
peak above 1 keV, interestingly at the same location
of the absorption features detected by the line scan (see Fig. \ref{Fig:Plot_RGS_all} 
and \ref{Fig:Plot_line_search}, and previously published in \citealt{Pinto2016nature}
for both RGS and EPIC spectra). 

\section{Discussion}
\label{sec:discussion}

\subsection{The NGC 1313 X-1 campaign}
\label{sec:discussion_campaign}


It is currently believed that the majority of ultraluminous X-ray sources (ULXs)
consists of stellar-mass compact objects accreting above the Eddington limit
by factors of a few up to hundreds, \textcolor{black}{particularly after the discoveries of pulsars
in some ULXs and mildly relativistic winds as predicted by models of super-Eddington accretion.
This research field is rather young and several problems are still to be solved regarding
the nature and launching mechanism(s) of the winds, their energy budget and effects
on the circumstellar medium and on the life of the binary system.}

We have been awarded 6 orbits targeting the NGC\,1313 galaxy 
with XMM-\textit{Newton} with the aim of achieving a better understanding of ULX winds.
Some of these observations have been taken simultaneously with 
\textit{NuSTAR} in order to improve the coverage of the hard X-ray band
and \textit{Chandra} to search for further narrow lines in the 2$-$7 keV energy band,
which is not covered by RGS and where EPIC has low energy resolution.
In this first paper we focus on the study of the wind in NGC 1313 X-1 and, in particular, 
its variability with the ULX spectral state based on the well-known spectral variability
\citep{Middleton2015a} and detection of emission and absorption lines 
\citep{Pinto2016nature}.
We notice that this campaign has already provided interesting results such as the first
discovery of pulsations in NGC 1313 ULX-2 \citep{Sathyaprakash2019a},
which added a new candidate to the presently small list of PULXs.

The study of the broadband spectral continuum is presented in a companion
paper \citep{Walton2020}. 
\textcolor{black}{The search for additional narrow spectral features at intermediate
energies (1$-$7 keV) with \textit{Chandra} is presented in Nowak et al. (in prep). 
The spectral-timing analysis and the application of Fourier techniques onto our 
new data is also described in another paper of this series \citep{Kara2020}
where we show the first discovery of a soft lag in NGC 1313 X-1
similarly to other two ULXs with wind detections: NGC 5408 X-1 and NGC 55 X-1.}

\textcolor{black}{Here we briefly note that our observations scheduled in pairs separated by 
long periods of 2-3 months enabled us to detect NGC 1313 X-1 at different fluxes
covering a smooth range of spectra} (see \textit{Swift}/XRT lightcurve in Fig.\,\ref{Fig:Plot_Swift_all}
and spectra Fig.\,\ref{Fig:Plot_RGS_all} and \ref{Fig:Plot_EPIC_all}).
The source shows a continuum behaviour typical of the subclass of ULXs with 
hard spectra at low fluxes and softer, broad disc-like, spectra at high fluxes in
according to the classification of \citet{Sutton2013}.

XMM-\textit{Newton} EPIC and RGS spectra from observations with similar flux levels 
and spectral shape have been \textcolor{black}{combined to compare the wind appearance in three different 
epochs and luminosity regimes (see Fig.\,\ref{Fig:Plot_RGS_all}).}
The three broadband spectra have been modelled with a multi-component spectral 
continuum model consisting of two modified-disc-blackbody components to reproduce the 
classical two broad peaks around 1 and 3$-$7 keV in the low state 
(see Table\,\ref{table:continuum}, Fig.\,\ref{Fig:Plot_RGS_all} and \ref{Fig:Plot_EPIC_all}) 
plus a comptonisation
component to fit the hard tail often found in ULXs (see, e.g., \citealt{Walton2018a}).

Our results are broadly consistent with those presented in the paper by \citet{Walton2020}
\textcolor{black}{to whom we refer for a more detailed analysis of the X-ray 
broadband continuum and of its variability}.
During the bright state the spectral curvature around 2 keV decreases 
(see Fig.\,\ref{Fig:Plot_RGS_all}) and the two thermal components are not well 
\textcolor{black}{distinguished} in our model (see Table\,\ref{table:continuum}) 
but this has no effects on the search for narrow spectral features. 


\subsection{The wind}
\label{sec:discussion_wind}

\subsubsection{Emission lines}
\label{sec:discussion_wind_emission}

Strong positive residuals are found in correspondence of the transition energies of several 
ions such as {\ovii}, {\oviii}, {\nex} and {\fexvii} in agreement with \citet{Pinto2016nature} and confirmed
in this work (see Fig.\,\ref{Fig:Plot_RGS_all_res} and \ref{Fig:Plot_line_search}).

Some emission lines clearly change, becoming stronger in the softer, bright, state
(see Fig.\,\ref{Fig:Plot_Gaussians}). This occurs for {\oviii} and several lines around
1 keV. \textcolor{black}{However, their equivalent widths are lower in the bright state
because the continuum around 1 keV is about 4 times higher than in the L-I hard state}. 
This broadly agrees 
with the results \citet{Middleton2015b} who showed that the residuals
around 1 keV weaken in the \textcolor{black}{bright} spectra. 
Hints of variability in both the absorption and emission
lines were already found in \citet{Pinto2016nature}, but this is the first time that we can
robustly confirm it. Photoionisation emission modelling further strengthens the variability 
of the emission lines with the spectral state and with the time (see Fig.\,\ref{Fig:emitting_gas}).
This is the first unambiguous evidence that the emission lines are intrinsic to the ULX. 

\begin{figure*}
  \includegraphics[width=0.8\columnwidth, angle=0]{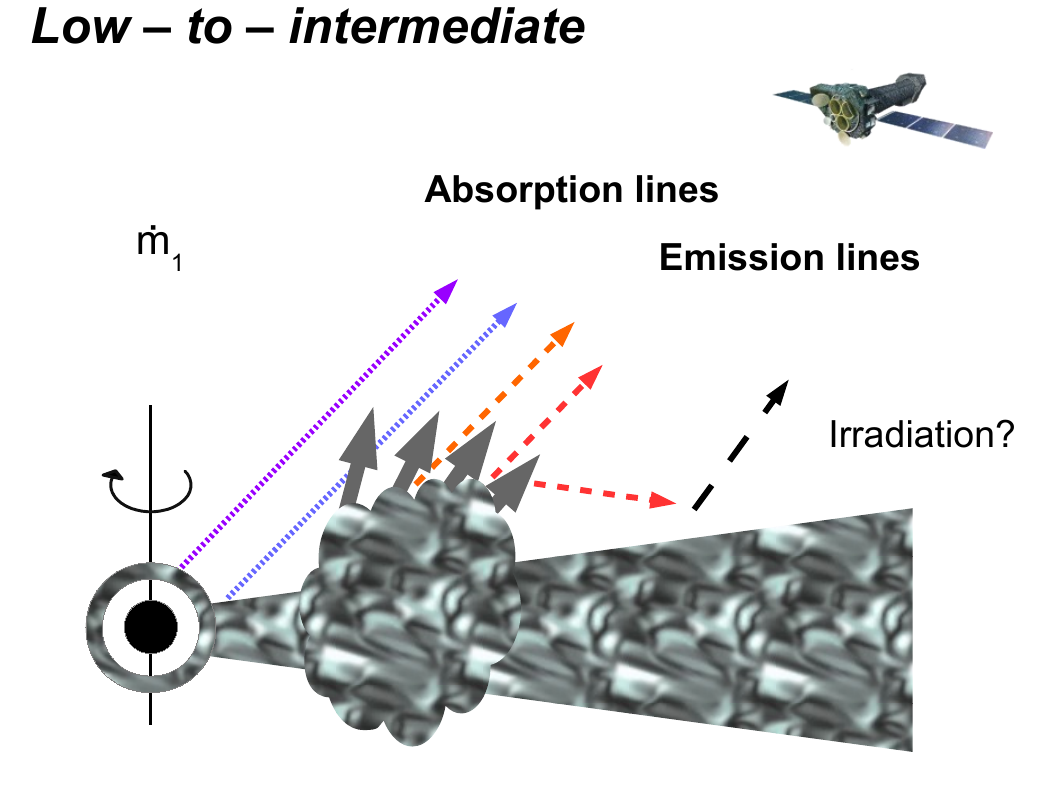}
  \hspace{1cm}
  \includegraphics[width=0.85\columnwidth, angle=0]{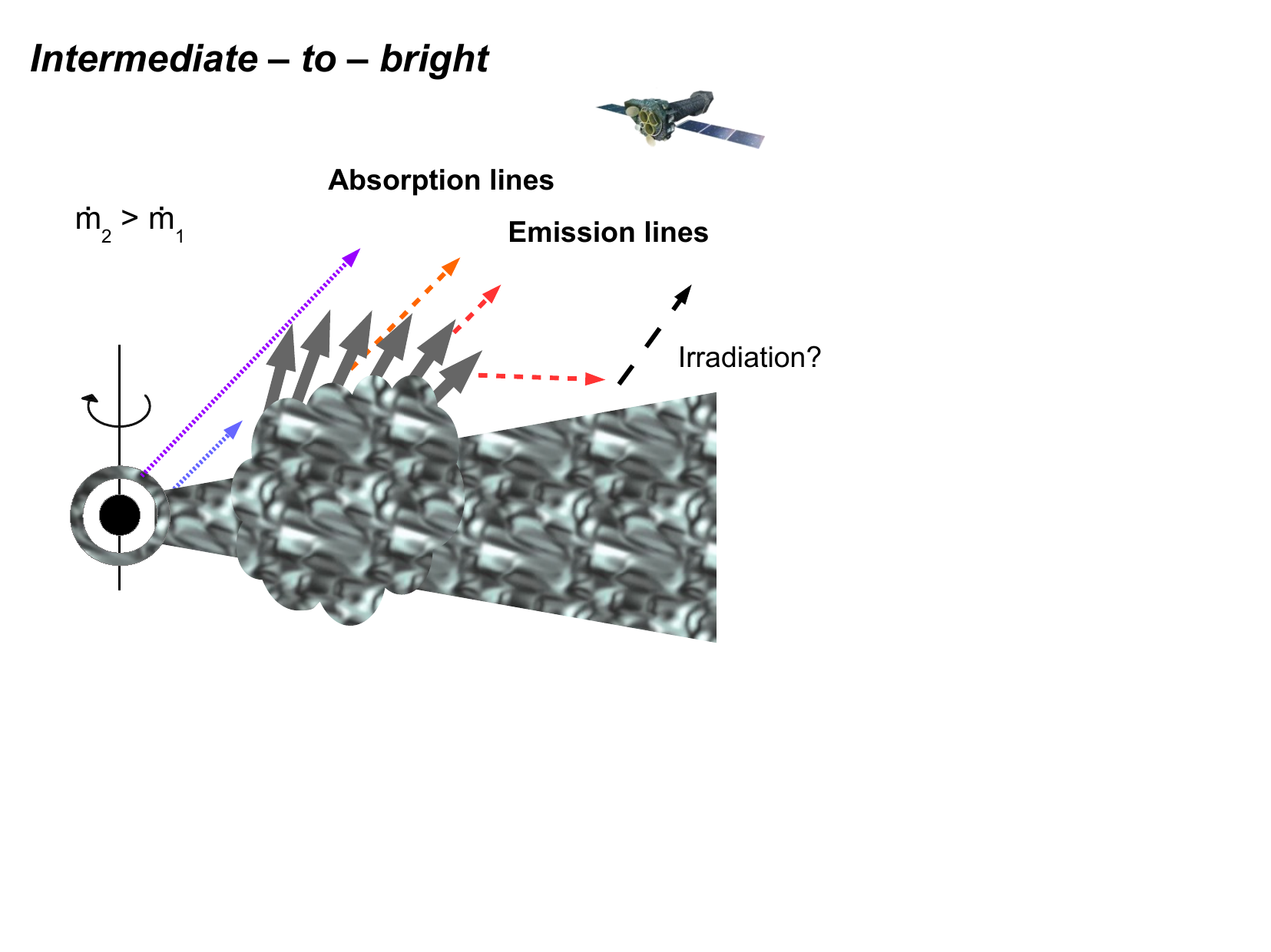}
  \vspace{-0.3cm} 
   \caption{Simplified scheme of a variable high mass accretion rate source.
                 The compact object is surrounded by an inner corona (or accretion column)
                 and a thick disc of matter from the companion star.
                \textit{Left panel:} Above the Eddington limit ($\dot{m}_1 > \dot{m}_E$)
                radiation pressure thickens the disc and launches a powerful wind from a region 
                around the spherisation radius ($R\sim25\,R_S$ for $v_{\rm esc}\sim0.2c$). 
                \textit{Right panel:} At even higher accretion rates ($\dot{m}_2 > \dot{m}_1 > \dot{m}_E$), 
                the source simultaneously brightens and softens, 
                the spherisation radius extends to larger radii and a cooler phase of the wind 
                is launched at larger radii ($\sim150\,R_S$ for $0.08c$).
                The emission lines, presumably coming from the outer irradiated disc,
                also become stronger and broader. These trends are observed in NGC 1313 X-1.}
   \label{Fig:Geometry}
  \vspace{-0.3cm} 
\end{figure*}

The emission lines are stronger, broader, and likely redshifted in the bright state
(see Fig.\,\ref{Fig:Plot_Gaussians_OVIIIvel}), which would be consistent with them
being emitted from the outer disc and progressively stronger at high accretion rates.
They resemble the flat-top profile with steep wings of Paschen lines in broad line regions
of AGN or disc winds of X-ray binaries such as GRO J1655-40 (\citealt{Soria2000}). 
In this case the emission lines may be produced near or outside the spherisation radius, in the irradiated 
part of the disc, with their strength proportional to accretion rate (see cartoon in Fig.\,\ref{Fig:Geometry}). 

The $g,r$ line ratios of the {\ovii} triplet suggest a high density hybrid gas 
($n_{\rm H} \sim 10^{10-12}$ cm$^{-3}$) where both recombination and collisional processes 
occur \citep{Porquet2000}. 
Models of collisionally-ionised emitting plasma 
provide similar results to those obtained using photoionisation equilibrium, but generally 
slightly worse fits (see Appendix\,\ref{sec:appendix} and Fig.\,\ref{Fig:emitting_gas_CIE}). 
Photoionisation is therefore preferred, but we cannot rule out some
contribution from collisions e.g. due to shock between the wind and the companion star
or the surrounding interstellar medium.

The luminosity of the line-emission component is remarkably high
($L_{\,0.3-10\,\rm keV}>10^{38}$\,erg/s), about 2-3 orders of magnitude higher 
than winds - including colliding winds - in classical supergiant X-ray binaries 
(hosting a neutron star accreting below the 
Eddington limit from a supergiant OB star, see e.g. \citealt{ElMellah2017}) and much higher 
than lines from accretion disc coronae of low-mass X-ray binaries 
(LMXBs, see, e.g., \citealt{Kallman2003}). \textcolor{black}{CCD spectra provide
comparable results (see, e.g., \citealt{Middleton2015b} and references therein).
Such luminosity is consistent with that measured by \citet{Wang2019} 
for the plasma producing blueshifted emission lines
in \textit{Chandra} and XMM-\textit{Newton} spectra of a transient ULX in NGC 6946}.

Interestingly, both the column densities of $10^{21-22}$ cm$^{-2}$ and
the volume densities ($n_{\rm H} \sim 10^{10-12}$ cm$^{-3}$)
measured for the line-emitting gas in NGC 1313 X-1 (assuming Solar abundances, 
\textcolor{black}{see Table\,\ref{table:improvements}}) 
are comparable to those measured in accretion disc coronae of some LMXBs 
(see, e.g., \citealt{Cottam2001}, \citealt{vanPeet2009}, \citealt{Iaria2013} and \citealt{Psaradaki2018}).
We note that we adopt a full solid angle for the emitting gas in order to speed up our fit in this paper. 
A lower solid angle of e.g. $\Omega/4\pi\sim0.1$, as found in the plasmas
around many X-ray binaries, would yield a larger column density in NGC 1313 X-1.
\textcolor{black}{We remark that our volume density estimates might be affected by 
uncertainties due to the low statistics above 20\,{\AA} and the possible contribution
from collisional ionisation. For a volume density $n_{\rm H} \sim 10^{12}$ cm$^{-3}$ and an 
ionisation parameter $\xi \sim 100$ erg/s cm, using Eq.\,(\ref{Eq:Eq_ionpar}), we obtain
a distance $R \sim 0.5$ U.A., which is comparable to the size of the roche lobe estimated 
for the pulsating ULX NGC 7793 P13 by \citet{Fuerst2018}.}

\subsubsection{Absorption lines}
\label{sec:discussion_wind_absorption}

In \citet{Pinto2016nature}, a highly significant ultrafast outflow was detected in NGC 1313 X-1
through the combination of the EPIC and RGS absorption features in the soft X-ray band.
Here we confirm the detection, even using only RGS below 2 keV 
(albeit at lower significance because of the more restrictive data selection used in this work; 
see Fig.\,\ref{Fig:Plot_line_search}, top left panel). 
The wind has remarkably changed from 2012 (and before)
to 2017 with the fast, $\sim0.2c$, and highly-ionised,
$\log \xi \sim 2.3$, component significantly weaker and a slower, $0.08c$,
and lower-ionisation, $\log \xi \sim 0.8$, phase now
also clearly detected (see Fig.\,\ref{Fig:absorbing_gas}).
The relative strengthening of such a `cooler' phase compared to 
the `hotter' one is more evident in the bright state.
The absorption lines of this new cooler phase appear broader 
(see Fig.\,\ref{Fig:Plot_line_search} \textcolor{black}{and Table\,\ref{table:improvements}}).

These findings are corroborated by the results of the principal component analysis
(see Fig.\,\ref{Fig:Plot_PCA}) which showed \textcolor{black}{new features} 
at intermediate energies that agree with the most relevant Lyman\,$\alpha$ lines
of Mg, Si, S and Ar all blueshifted by about $0.08c$.
The variations in the wind velocity and ionisation parameter may explain 
why its features are weak in the PCA. This technique commonly shows
stronger features in coherently-changing winds and at consistent velocities.

\subsection{A scenario of variability in the accretion rate}
\label{sec:discussion_wind_continuum_accretion}

Our results are consistent with a scenario in which a higher accretion rate 
has occurred during the middle and the end of 2017 and, possibly, 
during the previous year ({as suggested by the high variability and average flux
levels in the \textit{Swift} light curve} from 2016, see Fig.\,\ref{Fig:Plot_Swift_all}).
{We speculate that the recent higher accretion rate has caused} 
the brighter and softer state of the source, 
the appearance of a cooler and slower component of the wind, the strengthening of the emission lines
and the broadening of most emission and absorption lines.

\textcolor{black}{Regarding the emission lines, assuming that they are produced near the 
spherisation radius, an increase in the accretion rate would move the spherisation radius 
outwards, broadening the emitting region and the velocity pattern.
Otherwise, if they are emitted from the portion of the disc external to the spherisation radius,
then their broadening might be caused by the inner wind that increasingly shields the outer disc 
at higher $\dot{m}$, which lowers the ionisation of the disc and enables emission of (broader) lines 
from smaller radii, where previously the ions were fully ionised.}

Assuming that the velocities measured in our line of sight are proxies
for the (escape) launching velocities of the wind, we estimate an outer launching radius
($\sim150\,R_S$ for $0.08c$) than for the faster component detected previously
($\sim25\,R_S$ for $0.2c$). {The smaller launching radius and the higher velocity
are comparable to those estimated in the inner regions of the disc through GR-MHD 
simulations, while the slower and outer wind are similar to the outer region where 
the wind becomes clumpy (see e.g. \citealt{Takeuchi2013} and \citealt{Kobayashi2018}).}

The emission lines would be produced 
even further out (see Fig.\,\ref{Fig:Geometry}).
During the L-I state in September 2017 the imbalance between
the slow and fast phases seems milder than in the I-B state
\textcolor{black}{(see Fig.\,\ref{Fig:absorbing_gas} and Table\,\ref{table:improvements})}, 
maybe due to a lower accretion rate. 
The fast wind could be still hard to detect if the inner regions are obscured by the 
newly discovered cooler phase of the wind (as illustrated in Fig.\,\ref{Fig:Geometry},
right panel).
We note that the lower statistics of the new L-I state 
prevent us from making strong claims.
\textcolor{black}{Finally, the kinetic power of the two main wind components is comparable
($L_{\rm kin, \, Fast / Slow} \sim1$).
This is because the velocity of the fast component $v_{\rm \, Fast} \sim 2-3 \, v_{\rm \, Slow}$, 
the ionisation parameter $\xi_{\rm \, Fast} \sim 10-30 \, \xi_{\rm \, Slow}$, the ionising luminosity
(see Table\,\ref{table:improvements})
$L_{\rm ion, \, I-B} \sim 1.5 \, L_{\rm ion, \, L-I}$ (see e.g. Table\,\ref{table:continuum}) 
and the kinetic power goes as 
$L_{\rm kin} \propto L_{\rm ion}\,(v^3_{\rm out} / \xi)\,C_V\,\Omega$.
Here we assume the same covering factor and opening angle of the wind components,
which can of course differ.}


In summary, the general picture may be that the X-ray continuum comes from the inner 
(super-critical) disc, the line absorption comes from the wind launched inside the spherisation radius 
and the line emission comes from the thinner, slower wind launched from the irradiated 
standard-disc surface outside the spherisation radius.
This is consistent with the fact that blueshifted emission lines are seen in NGC 55 ULX
(see \citealt{Pinto2017}), a spectral hybrid between soft ULXs and supersoft ultraluminous 
sources (see, e.g., \citealt{Feng2016}) that is believed to be viewed at a larger inclination 
than NGC 1313 X-1.
\textcolor{black}{Further, in-depth, work is required to confirm this scenario and indeed
several forthcoming papers are being produced focusing on the broadband variability spectral
variability on different time scales (e.g. \citealt{Walton2020} and \citealt{Kara2020}).}

\subsection{\textcolor{black}{{Future studies of ULX winds with \textit{ATHENA}}}}
\label{sec:discussion_missions}

\textcolor{black}{Current X-ray observatories such as 
XMM-\textit{Newton} and \textit{Chandra} 
have opened up new research channels to understand ULXs and, more in general,
super-Eddington accretion. 
However, their high-energy-resolution spectrometers have low effective area,
requiring exposure times of 1-2 days to achieve high statistics.
This prevent us from tracking the wind variability down to time scales
as short as 1 ks (or even less), which are similar to the time lags found in 
some ULXs including NGC 1313 X-1 \citep{Kara2020} 
and those shown by the PCA (Sect.\,\ref{sec:timing_analysis}).
The study of the winds behaviour at such time scales could place 
stronger constraints on their nature and structure.
The future X-ray mission \textit{XRISM} \citep{Guainazzi2018} 
will enable to detect and resolve features from hotter wind phases 
in the hard X-ray band ($\sim2-10$\,keV), 
providing complementary information to RGS but still requiring 
days of observations (see \citealt{Pinto2019a} for a \textit{XRISM} simulation
of NGC 1313 X-1).}


\textcolor{black}{X-ray astronomy will dramatically change in the 2030s after the launch of ATHENA 
(\citealt{Nandra2013}). Its X-ray Integral Field Unit (X-IFU), in particular, will be 
the first X-ray microcalorimeter to provide both high spatial (5'') and spectral (2.5 eV)
resolution with an effective area an order of magnitude higher
than any current high-spectral-resolution X-ray detector, which gives
an improvement of almost 2 orders of magnitude in the detection
of narrow features in the canonical 0.3$-$10 keV X-ray energy band
(see e.g. \citealt{Barret2018}).}

\begin{figure}
  \includegraphics[width=0.95\columnwidth, angle=0]{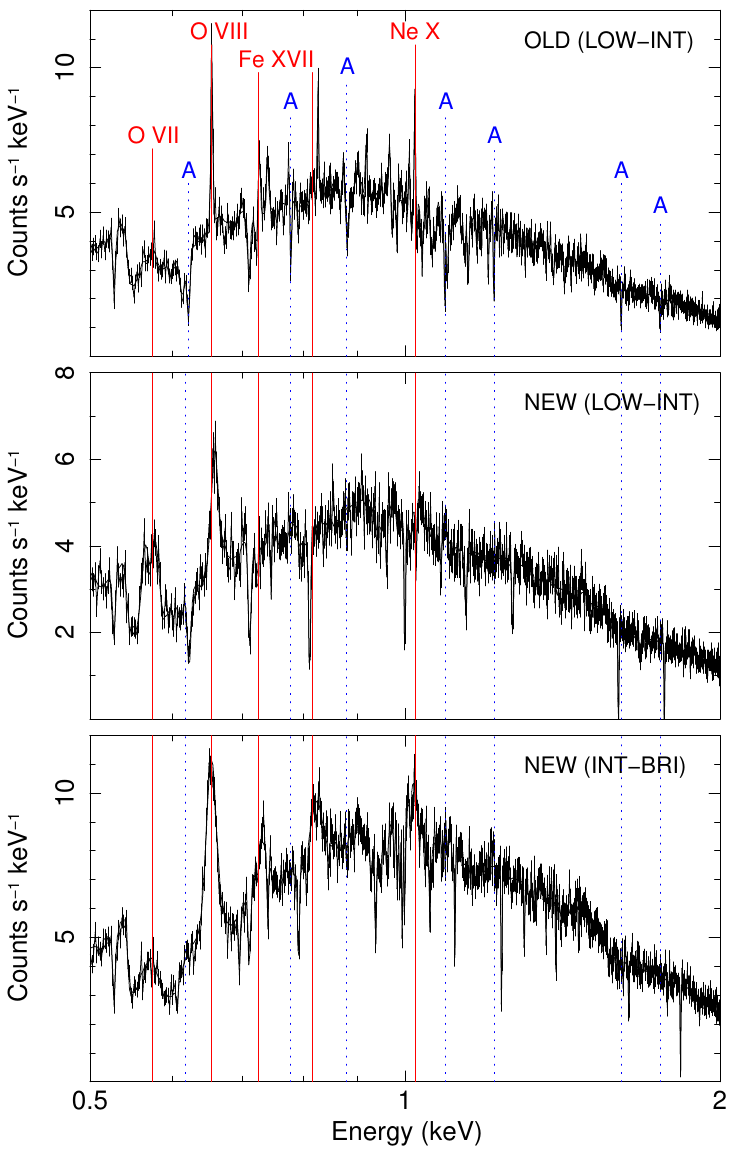}
   \vspace{-0.3cm}
   \caption{\textcolor{black}{ATHENA\,/\,X-IFU 
   simulations of the ULX NGC 1313 X-1 for the 
                 three spectra with overlaid the best-fit multiphase wind model.
         The wind model consists of photoionised emitting plasma and a 
         two-phase absorber (see Table\,\ref{table:improvements}
         and Fig.\,\ref{Fig:rgs_spectrum_fit}).
                 Cosmic X-ray background and particle background are accounted for.
                 \textcolor{black}{The X-IFU response matrix is dated 10/2018 with 1.4\,m$^2$ 
                 effective area at 1 keV.}
                 Note how the lines strongly differ in these spectra.}}
   \label{Fig:plot_ATHENA}
   \vspace{-0.3cm}
\end{figure}

\textcolor{black}{We simulate three X-IFU spectra of NGC 1313 X-1 using the best-fit wind models 
of the three RGS spectra
(see Table\,\ref{table:improvements} and Fig.\,\ref{Fig:rgs_spectrum_fit}) 
and adopting an X-IFU exposure time of 25\,ks each. 
The expected results are extremely promising with the lines obviously changing
between the states, while in RGS spectra - without a careful analysis - 
might appear comparable (Fig.\,\ref{Fig:plot_ATHENA}).
A short snapshot of 1\,ks with the X-IFU will still provide a high signal-to-noise 
spectrum with several emission and absorption lines detected 
at $5\,\sigma$ each.
This will enable to study the wind at the variability time scales 
probed by the PCA (Sect.\,\ref{sec:timing_analysis})
and by ULX soft lags (few\,10-to-1000\,s).
Moreover, the X-IFU will detect winds in ULXs at much larger distance,
boosting the explored volume of ULXs and determining 
wind properties at much higher accuracy.
The comparison between their properties with those of the interstellar cavities 
around many ULXs will determine the role ULXs feedback, particularly in the early Universe.}

\section{Conclusions}
\label{sec:conclusion}

Most ULXs are believed to be powered by super-Eddington accreting neutron stars and black holes.
If the compact object is not a neutron star with a magnetar-like magnetic field,
the disc is expected to thicken and launch powerful winds through radiation pressure
or magnetic fields might launch winds themselves.
Evidence of winds has been found in the ULXs through the high-resolution
spectrometers onboard XMM-\textit{Newton}. In order to \textcolor{black}{understand the wind 
geometry, its relationship with the accretion rate and the source state}, 
we undertook a large observing campaign with 
XMM-\textit{Newton} to study the NGC 1313 (UL)X-1,
which exhibits strong rest-frame emission and blueshifted absorption lines.

The new observations show remarkable wind variability, with the previously known fast 
component ($\sim0.2c$) to have weakened, and a new, cooler and slower ($\sim0.08c$), wind 
component appearing. Significant variability is found in the highly significant emission lines,
showing for the first time that they are most likely produced in the accretion disc 
\textcolor{black}{or the outer wind}.
We describe the simultaneous change of the wind and the spectral continuum
in terms of variability in the super-Eddington accretion rate. The geometry that 
can describe the observables and agrees with the theoretical models of super-Eddington
accretion discs, implies that the continuum is emitted within the inner super-critical regions,
the line absorption is produced by the wind launched near the spherisation radius
and the emission lines come from a slower wind component in the outer, likely irradiated, disc.

\appendix

\section{Technical details}
\label{sec:appendix}
 
\textcolor{black}{In this section we 
put the plots and tables that are excluded from the main body of the paper
to facilitate the reading. }

\begin{figure*}
  \includegraphics[width=1\columnwidth, angle=0]{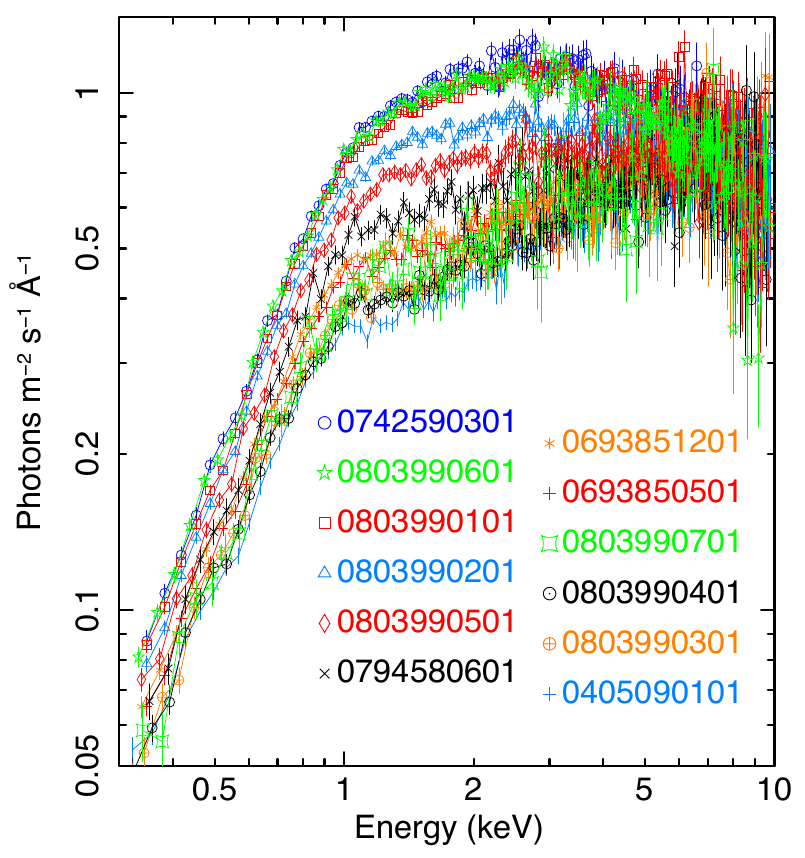}
  \includegraphics[width=1\columnwidth, angle=0]{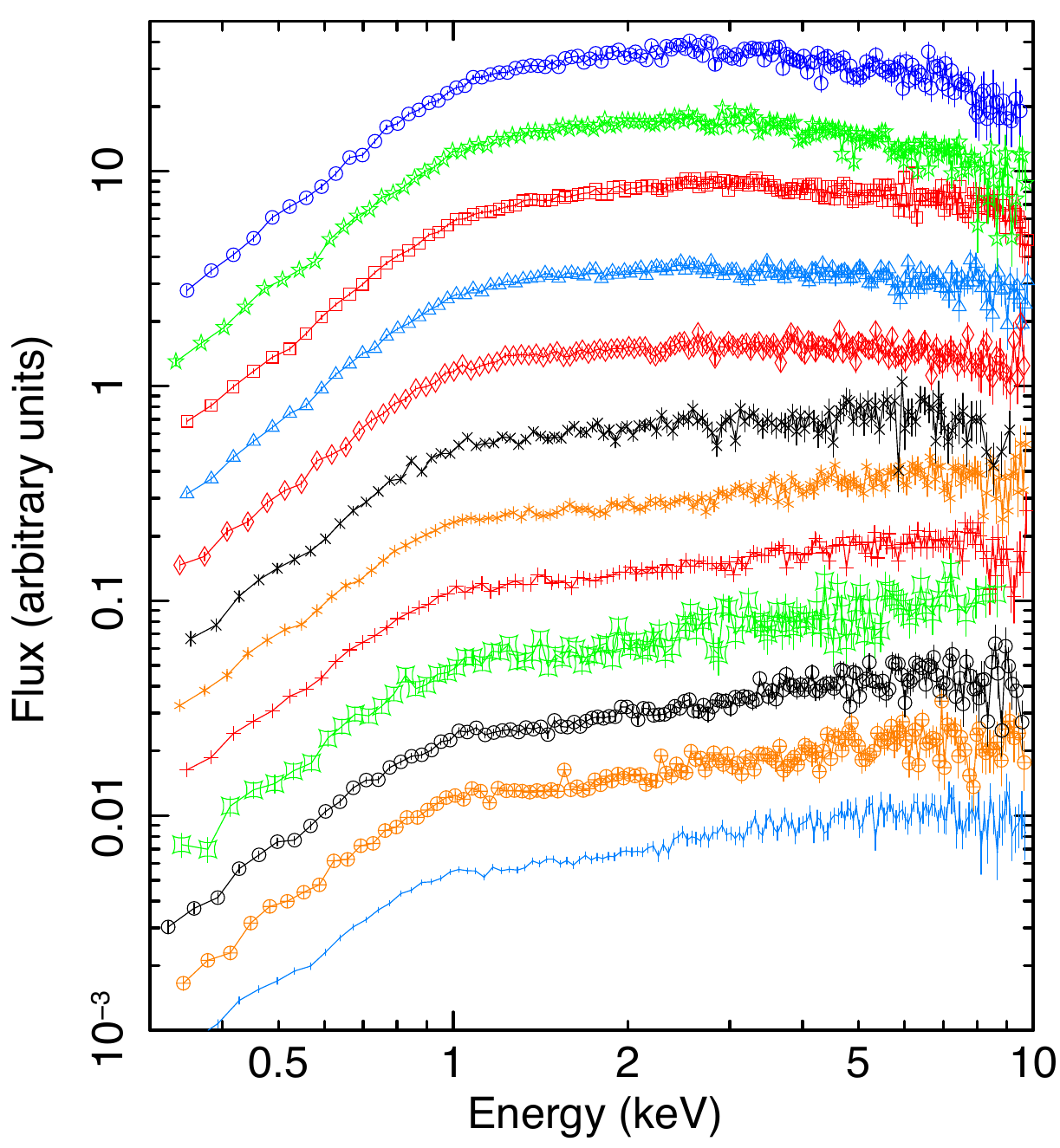}
   \vspace{-0.3cm}
   \caption{\textit{Left panel:} XMM-\textit{Newton}/EPIC-pn spectra for both archival and new (080399)
   observations. The spectra cover a continuous distribution of fluxes from low 
   (ultraluminous-hard) to the bright (broadened-disc) states. 
   Most flux variations occur in the soft-intermediate X-ray energy band, 
   while the spectra overlap beyond 5 keV. \textit{Right panel:}
   same data and color code as shown in the left panel, but here
   the spectra have been shifted along the Y-axis for displaying purposes.}
   \label{Fig:Plot_EPIC_all}
   \vspace{-0.3cm}
\end{figure*}

\textcolor{black}{Fig.\,\ref{Fig:Plot_EPIC_all} (left panel) shows the EPIC-pn spectra extracted for the 12 
individual exposures. In the right panel the spectra are in the same units,
but have been shifted in the Y-axis, multiplying for a constant for display purposes.}
  
\begin{figure*}
  \includegraphics[width=1\columnwidth, angle=0]{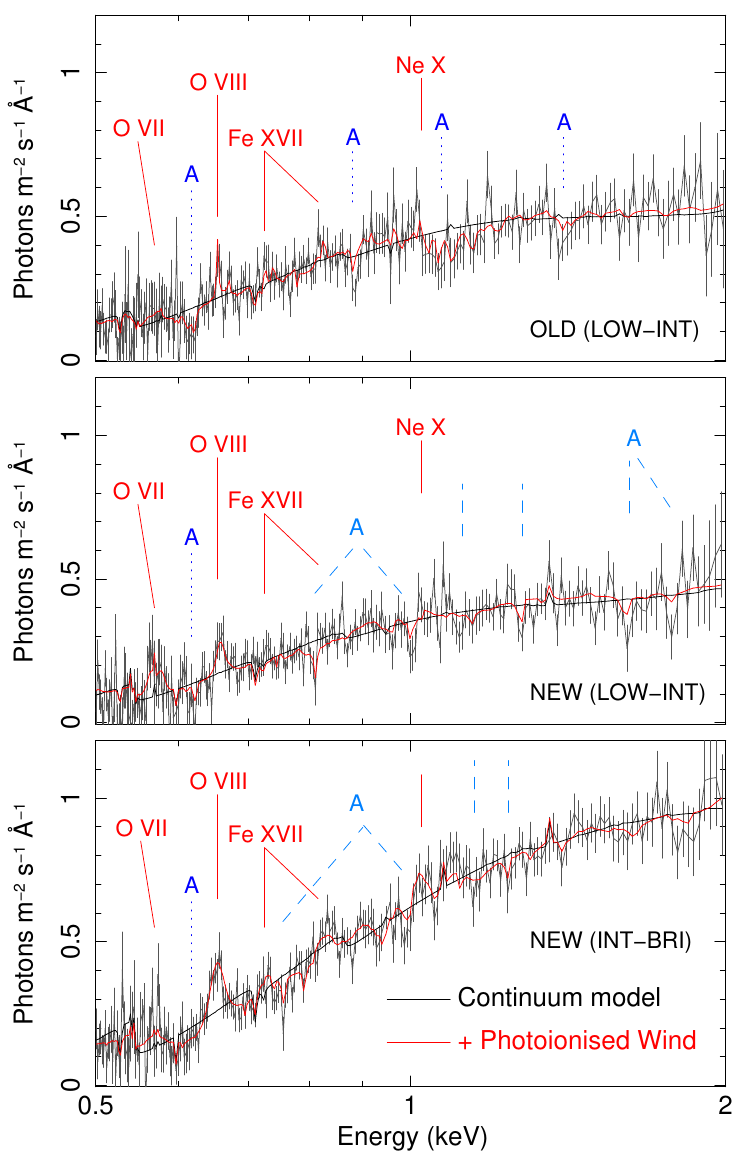}
  \includegraphics[width=0.889\columnwidth, angle=0]{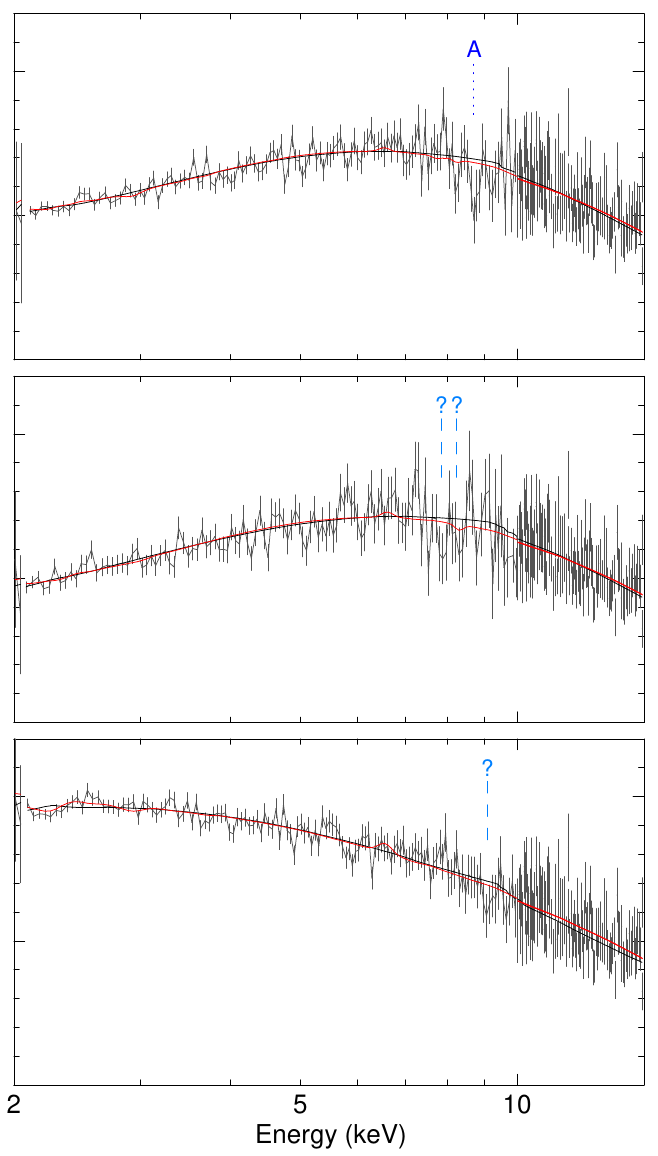}
   \vspace{-0.3cm}
   \caption{\textcolor{black}{RGS (left), EPIC-pn and FPM\,A,B(right) spectra 
         with overlaid the best-fit multiphase wind model (red line) 
         and the broadband continuum model (black line).
         The wind model consists of a two-phase photoionised absorber 
         and a photoionised emission component
         (see Table\,\ref{table:improvements}).
         Data and line labels are the same as in Fig.\,\ref{Fig:Plot_RGS_all}.}} 
            \label{Fig:rgs_spectrum_fit}
\vspace{-0.3cm}
\end{figure*}

\textcolor{black}{The best-fit wind models with the three main emission and absorption
components are shown in Fig.\,\ref{Fig:rgs_spectrum_fit} 
with their relevant parameters detailed in Table\,\ref{table:improvements}.
We also quote the spectral fit improvement $\Delta C$ and the significance $\sigma$ for all the
components.
We report them in the case where each component was fit alone with the continuum as for 
the model scans ($\Delta C^a$ and $\sigma^a$) and also where all emission and absorption components
where included in the model ($\Delta C^e$ and $\sigma^e$). The latter represents a test for independence
among the several components. Here their individual significances decrease but each of them still provides
substantial improvements to the spectral fits, some of which well beyond 3 and 4\,$\sigma$.}
 
\begin{figure}
  \includegraphics[width=1\columnwidth, angle=0]{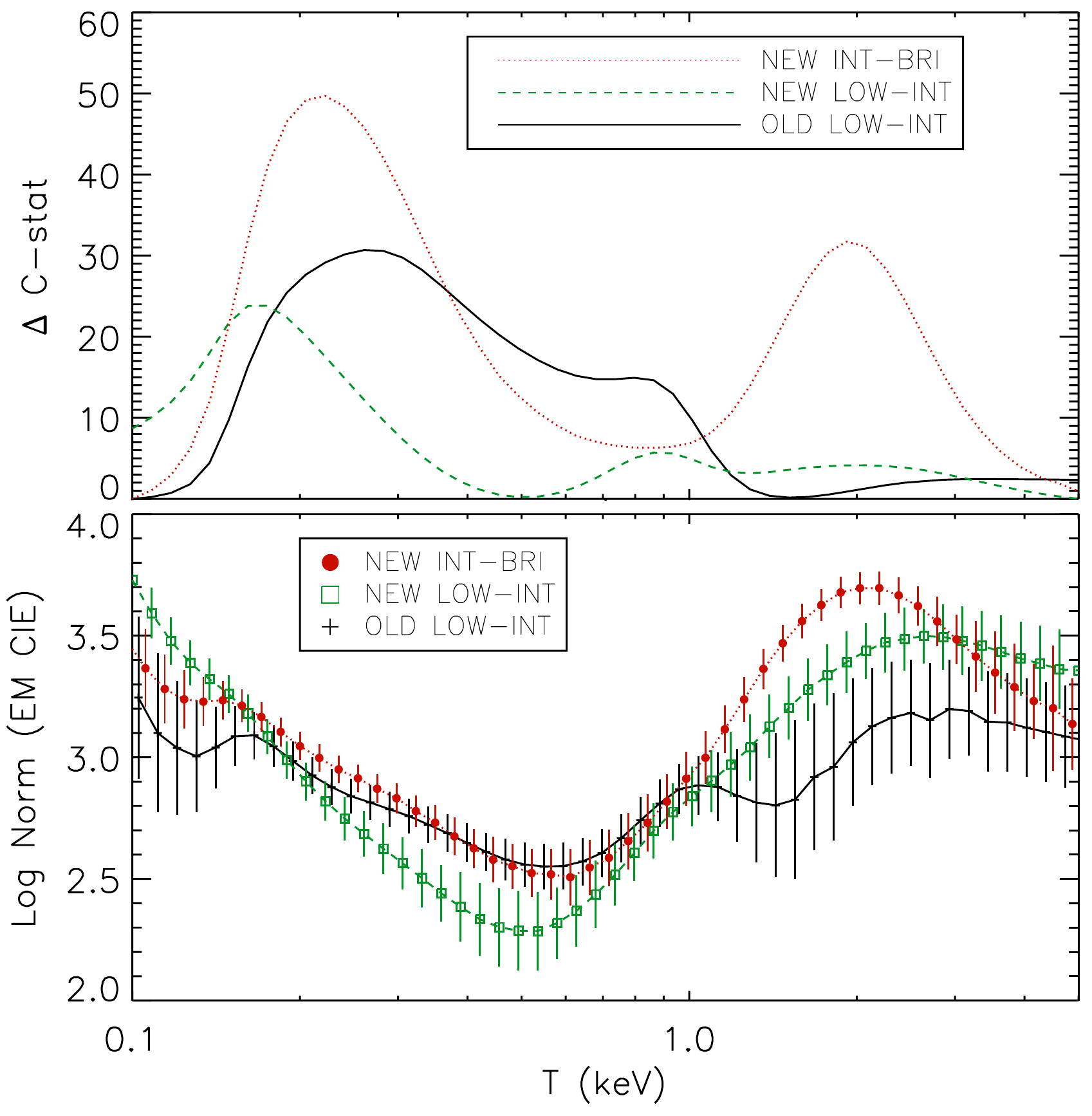}
   \vspace{-0.5cm}
   \caption{Wind search through scans of collisional ionisation emission models
   for the archival low-to-intermediate (solid black line),
   the new low-to-intermediate (dashed green line) and the new intermediate-to-bright
   spectra (dotted red line); see also Sect.\,\ref{sec:emitting_gas}. 
   The X-axis shows the plasma temperature, $kT$ (slightly shifted for displaying purposes). 
   The top panel shows $\Delta C$-stat fit improvement for each $kT$
   and the bottom panel is the corresponding emission measure
   (Log $n_e n_{\rm H} V \, [ 10^{58}$ cm$^{-3}$]).}
   \label{Fig:emitting_gas_CIE}
   \vspace{-0.4cm}
\end{figure}

\textcolor{black}{In Fig.\,\ref{Fig:emitting_gas_CIE} we show the search for a shocked wind performed by scanning 
an emission model of gas in collisional equilibrium ($cie$ model in {\scriptsize{SPEX}}) 
over the three flux-/time-resolved spectra: 
archival low-to-intermediate state (solid black line), the new low-to-intermediate state (dashed green line) 
and the new intermediate-to-bright state (dotted red line); see also Sect.\,\ref{sec:emitting_gas}. 
For a comparison with a similar scan using photoionisation emission 
models see Fig.\,\ref{Fig:emitting_gas}. Collisional-ionisation equilibrium provides generally
worse fits than photoionisation except for the archival data.}

\begin{table}
\caption{\textcolor{black}{NGC 1313 X-1: main wind components.}}  
\label{table:improvements}             
\renewcommand{\arraystretch}{1.}      
 \small\addtolength{\tabcolsep}{-2.5pt}
 
\scalebox{0.95}{%
\hskip-0.5cm\begin{tabular}{@{}l l l l l l l l}     
\hline  
Phase                     & Parameter                       &  OLD L-I                   &  NEW L-I            &  NEW I-B            \\  
\hline                                                                                         
\multirow{4}{*}{EMI}     &   $N_{\rm H}$            & $3.0 \pm 0.8$          & $1.4 \pm 0.3$      & $3.5 \pm 1.3$     \\
\multirow{5}{*}{(rest)}   &   $\log \xi$                  & $2.4 \pm 0.1$          & $1.1 \pm 0.1$       & $2.2 \pm 0.2$     \\
                             &   $v_{\rm LOS}$ (km/s)      & $100\pm300$          & $2000\pm700$   & $-550\pm350$   \\
                                    &   $v_{\sigma}$  (km/s) & $200\pm 200$        & $1600\pm 400$    & $2100\pm 500$   \\
                                     &   $\Delta C^a(^e)$      & 25(21)                     & 40(35)                 & 55(50)                 \\
                                     &   $\sigma^a(^e)$        & 3.5(3.1)                   & $>4$($>4$)         & $>4$($>4$)         \\
\hline                                                                                         
\multirow{5}{*}{ABS\,1} &  $N_{\rm H}$             & $0.3 \pm 0.1$           &  $0.4 \pm 0.1$      &  $0.7 \pm 0.1$     \\
\multirow{5}{*}{(slow)}   &  $\log \xi$                   & $0.8 \pm 0.3$          &  $0.7 \pm 0.2$      &  $0.9 \pm 0.2$     \\
                                 &   $v_{\rm LOS}$ ($c$)  & $0.081\pm0.004$      & $0.084\pm0.003$  & $0.084\pm0.004$ \\
                                 &   $v_{\sigma}$  (km/s) & $700\pm400$             &  $800\pm400$       &  $3500\pm1500$  \\
                                     &   $\Delta C^a(^e)$     & 17(10)                       & 25(21)                   & 38(30)                   \\
                                     &   $\sigma^a(^e)$       & 2.6(2.0)                     & 3.5(3.2)                 & $>4$($>4$)          \\
\hline                                                                                         
\multirow{5}{*}{ABS\,2} &  $N_{\rm H}$              & $4.1 \pm 1.5$          & $0.7 \pm 0.2$        &  $0.6 \pm 0.2$       \\
\multirow{5}{*}{(fast)}    &  $\log \xi$                   & $2.3 \pm 0.1$         & $1.8 \pm 0.1$         &  $1.8 \pm 0.1$       \\
                                  &   $v_{\rm LOS}$  ($c$)  & $0.191\pm0.002$   & $0.240\pm0.002$   &  $0.210\pm0.002$ \\
                                  &   $v_{\sigma}$  (km/s) & $100\pm100$           & $420\pm260$         &   $120\pm120$      \\
                                     &   $\Delta C^a(^e)$      & 26(21)                     & 18(15)                    &  19(13)                   \\
                                     &   $\sigma^a(^e)$        & 3.7(3.1)                   & 2.9(2.5)                   &  2.9(2.3)                \\
\hline                                                                                                                
\end{tabular}}

\vspace{0.1cm}

\textcolor{black}{Best-fit wind components: photoionised emission (EMI) and absorption (ABS, see also 
Fig.\,\ref{Fig:absorbing_gas} and Fig.\,\ref{Fig:emitting_gas}).
The ionisation parameters are in log ($\xi$, erg cm s$^{-1}$) and the column densities, 
$N_{\rm H}$, in $10^{21} {\rm cm}^{-2}$. The line-of-sight velocities, $v_{\rm LOS}$,
are in units of light speed $c$ for ABS\,1,2 and km/s for EMI
(assumed positive for blueshift and motion towards the observed); 
the line widths, $v_{\sigma}$, are in km s$^{-1}$.
The spectra are defined in Sect.\,\ref{sec:fluxed_spectra}.
The $\Delta C^a(^e)$ refer to the $\Delta C$-statistics of each component computed 
when the component is the only one in model (a) or when the other two are included (e).
The same applies to the detection significances, $\sigma^a(^e)$, which are evaluated
with Monte Carlo simulations. }
\end{table}

\section*{Acknowledgments}

This work is based on observations obtained with XMM-\textit{Newton}, an
ESA science mission funded by ESA Member States and USA (NASA).
CP, MP and FF are supported by European Space Agency (ESA) Research Fellowships.
We acknowledge support from STFC and ERC Advanced Grant Feedback 340442
and the European Union's Horizon 2020 Programme 
under the AHEAD AO5 project (grant agreement n. 654215).
We also thank J. de Plaa and M. Mehdipour for support in optimising 
{\scriptsize{SPEX}} and I. Psaradaki for help with {\scriptsize{PYTHON}}.
We thank the XMM-\textit{Newton} SOC and, in particular, Jan-Uwe Ness
for excellent support in optimising our observing campaign.
{We finally acknowledge the anonymous referee for useful 
comments that improved the clarity of the paper.}

\bibliographystyle{mn2e}
\bibliography{bibliografia} 

\begin{thebibliography}{}

\bibitem[\protect\citeauthoryear{{Bachetti}, {Harrison}, {Walton},
  {Grefenstette}, {Chakrabarty} et~al.,}{{Bachetti}
  et~al.}{2014}]{Bachetti2014}
{Bachetti} M.,  {Harrison} F.~A.,  {Walton} D.~J.,  {Grefenstette} B.~W.,
  {Chakrabarty} D.,    et~al., 2014, \nat, 514, 202

\bibitem[\protect\citeauthoryear{{Bachetti}, {Rana}, {Walton}, {Barret},
  {Harrison} et~al.,}{{Bachetti} et~al.}{2013}]{Bachetti2013}
{Bachetti} M.,  {Rana} V.,  {Walton} D.~J.,  {Barret} D.,  {Harrison} F.~A.,
  et~al., 2013, \apj, 778, 163

\bibitem[\protect\citeauthoryear{{Barret}, {Lam Trong}, {den Herder}, {Piro}
  et~al.,}{{Barret} et~al.}{2018}]{Barret2018}
{Barret} D.,  {Lam Trong} T.,  {den Herder} J.-W.,  {Piro} L.,    et~al., 2018,
  in SPIE Vol. 10699 of Society of Photo-Optical Instrumentation Engineers
  (SPIE) Conference Series, {The ATHENA X-ray Integral Field Unit (X-IFU)}.
p. 106991G

\bibitem[\protect\citeauthoryear{{Brightman}, {Harrison}, {Bachetti}
  et~al.,}{{Brightman} et~al.}{2019}]{Brightman2019}
{Brightman} M.,  {Harrison} F.~A.,  {Bachetti} M.,    et~al., 2019, \apj, 873,
  115

\bibitem[\protect\citeauthoryear{{Brightman}, {Harrison}, {F{\"u}rst},
  {Middleton}, {Walton} et~al.,}{{Brightman} et~al.}{2018}]{Brightman2018}
{Brightman} M.,  {Harrison} F.~A.,  {F{\"u}rst} F.,  {Middleton} M.~J.,
  {Walton} D.~J.,    et~al., 2018, Nature Astronomy, 2, 312

\bibitem[\protect\citeauthoryear{{Carpano}, {Haberl}, {Maitra} \&
  {Vasilopoulos}}{{Carpano} et~al.}{2018}]{Carpano2018}
{Carpano} S.,  {Haberl} F.,  {Maitra} C.,    {Vasilopoulos} G.,  2018, \mnras,
  476, L45

\bibitem[\protect\citeauthoryear{{Cash}}{{Cash}}{1979}]{Cash1979}
{Cash} W.,  1979, \apj, 228, 939

\bibitem[\protect\citeauthoryear{{Cottam}, {Sako}, {Kahn}, {Paerels} \&
  {Liedahl}}{{Cottam} et~al.}{2001}]{Cottam2001}
{Cottam} J.,  {Sako} M.,  {Kahn} S.~M.,  {Paerels} F.,    {Liedahl} D.~A.,
  2001, \apjl, 557, L101

\bibitem[\protect\citeauthoryear{{den Herder}, {Brinkman}, {Kahn} et~al.,}{{den
  Herder} et~al.}{2001}]{denherder2001}
{den Herder} J.~W.,  {Brinkman} A.~C.,  {Kahn} S.~M.,    et~al., 2001, A\&A,
  365, L7

\bibitem[\protect\citeauthoryear{{El Mellah} \& {Casse}}{{El Mellah} \&
  {Casse}}{2017}]{ElMellah2017}
{El Mellah} I.,  {Casse} F.,  2017, \mnras, 467, 2585

\bibitem[\protect\citeauthoryear{{Evans}, {Beardmore}, {Page}, {Osborne},
  {O'Brien} et~al.,}{{Evans} et~al.}{2009}]{Evans2009}
{Evans} P.~A.,  {Beardmore} A.~P.,  {Page} K.~L.,  {Osborne} J.~P.,  {O'Brien}
  P.~T.,    et~al., 2009, \mnras, 397, 1177

\bibitem[\protect\citeauthoryear{{Fabrika}}{{Fabrika}}{2004}]{Fabrika2004}
{Fabrika} S.,  2004, Astrophysics and Space Physics Reviews, 12, 1

\bibitem[\protect\citeauthoryear{{Farrell}, {Webb}, {Barret}, {Godet} \&
  {Rodrigues}}{{Farrell} et~al.}{2009}]{Farrell2009}
{Farrell} S.~A.,  {Webb} N.~A.,  {Barret} D.,  {Godet} O.,    {Rodrigues}
  J.~M.,  2009, \nat, 460, 73

\bibitem[\protect\citeauthoryear{{Feng}, {Tao}, {Kaaret} \& {Gris{\'e}}}{{Feng}
  et~al.}{2016}]{Feng2016}
{Feng} H.,  {Tao} L.,  {Kaaret} P.,    {Gris{\'e}} F.,  2016, \apj, 831, 117

\bibitem[\protect\citeauthoryear{{Francis} \& {Wills}}{{Francis} \&
  {Wills}}{1999}]{Francis1999}
{Francis} P.~J.,  {Wills} B.~J.,  1999, in {Ferland} G.,  {Baldwin} J.,  eds,
  Quasars and Cosmology Vol.~162 of Astronomical Society of the Pacific
  Conference Series, {Introduction to Principal Components Analysis}.
p.~363

\bibitem[\protect\citeauthoryear{{F{\"u}rst}, {Walton}, {Harrison}
  et~al.,}{{F{\"u}rst} et~al.}{2016}]{Fuerst2016}
{F{\"u}rst} F.,  {Walton} D.~J.,  {Harrison} F.~A.,    et~al., 2016, \apjl,
  831, L14

\bibitem[\protect\citeauthoryear{{F{\"u}rst}, {Walton}, {Heida}, {Harrison},
  {Barret} et~al.,}{{F{\"u}rst} et~al.}{2018}]{Fuerst2018}
{F{\"u}rst} F.,  {Walton} D.~J.,  {Heida} M.,  {Harrison} F.~A.,  {Barret} D.,
    et~al., 2018, \aap, 616, A186

\bibitem[\protect\citeauthoryear{{Gehrels}, {Chincarini}, {Giommi}, {Mason},
  {Nousek} et~al.,}{{Gehrels} et~al.}{2004}]{Gehrels2004}
{Gehrels} N.,  {Chincarini} G.,  {Giommi} P.,  {Mason} K.~O.,  {Nousek} J.~A.,
    et~al., 2004, \apj, 611, 1005

\bibitem[\protect\citeauthoryear{{Gladstone}, {Roberts} \& {Done}}{{Gladstone}
  et~al.}{2009}]{Gladstone2009}
{Gladstone} J.~C.,  {Roberts} T.~P.,    {Done} C.,  2009, \mnras, 397, 1836

\bibitem[\protect\citeauthoryear{{Greene} \& {Ho}}{{Greene} \&
  {Ho}}{2007}]{Greene2007}
{Greene} J.~E.,  {Ho} L.~C.,  2007, \apj, 656, 84

\bibitem[\protect\citeauthoryear{{Guainazzi} \& {Tashiro}}{{Guainazzi} \&
  {Tashiro}}{2018}]{Guainazzi2018}
{Guainazzi} M.,  {Tashiro} M.~S.,  2018, ArXiv e-prints

\bibitem[\protect\citeauthoryear{{Harrison}, {Craig}, {Christensen}, {Hailey},
  {Zhang} et~al.,}{{Harrison} et~al.}{2013}]{Harrison2013}
{Harrison} F.~A.,  {Craig} W.~W.,  {Christensen} F.~E.,  {Hailey} C.~J.,
  {Zhang} W.~W.,    et~al., 2013, \apj, 770, 103

\bibitem[\protect\citeauthoryear{{Iaria}, {Di Salvo}, {D'A{\`i}}, {Burderi},
  {Mineo}, {Riggio}, {Papitto} \& {Robba}}{{Iaria} et~al.}{2013}]{Iaria2013}
{Iaria} R.,  {Di Salvo} T.,  {D'A{\`i}} A.,  {Burderi} L.,  {Mineo} T.,
  {Riggio} A.,  {Papitto} A.,    {Robba} N.~R.,  2013, A\&A, 549, A33

\bibitem[\protect\citeauthoryear{{Israel}, {Belfiore}, {Stella}, {Esposito},
  {Casella} et~al.,}{{Israel} et~al.}{2017a}]{Israel2017a}
{Israel} G.~L.,  {Belfiore} A.,  {Stella} L.,  {Esposito} P.,  {Casella} P.,
  et~al., 2017a, Science, 355, 817

\bibitem[\protect\citeauthoryear{{Israel}, {Papitto}, {Esposito}, {Stella},
  {Zampieri} et~al.,}{{Israel} et~al.}{2017b}]{Israel2017b}
{Israel} G.~L.,  {Papitto} A.,  {Esposito} P.,  {Stella} L.,  {Zampieri} L.,
  et~al., 2017b, \mnras, 466, L48

\bibitem[\protect\citeauthoryear{{Kaaret}, {Feng} \& {Roberts}}{{Kaaret}
  et~al.}{2017}]{Kaaret2017}
{Kaaret} P.,  {Feng} H.,    {Roberts} T.~P.,  2017, \araa, 55, 303

\bibitem[\protect\citeauthoryear{{Kaaret}, {Simet} \& {Lang}}{{Kaaret}
  et~al.}{2006}]{Kaaret2006}
{Kaaret} P.,  {Simet} M.~G.,    {Lang} C.~C.,  2006, \apj, 646, 174

\bibitem[\protect\citeauthoryear{{Kaastra}, {Mewe} \&
  {Nieuwenhuijzen}}{{Kaastra} et~al.}{1996}]{kaastraspex}
{Kaastra} J.~S.,  {Mewe} R.,    {Nieuwenhuijzen} H.,  1996, in {K.~Yamashita \&
  T.~Watanabe} ed., UV and X-ray Spectroscopy of Astrophysical and Laboratory
  Plasmas {SPEX: a new code for spectral analysis of X {\&} UV spectra.}.
p.~411

\bibitem[\protect\citeauthoryear{{Kallman}, {Angelini}, {Boroson} \&
  {Cottam}}{{Kallman} et~al.}{2003}]{Kallman2003}
{Kallman} T.~R.,  {Angelini} L.,  {Boroson} B.,    {Cottam} J.,  2003, \apj,
  583, 861

\bibitem[\protect\citeauthoryear{{Kara}, {Pinto}, {Walton}, {Alston},
  {Bachetti} et~al.,}{{Kara} et~al.}{2020}]{Kara2020}
{Kara} E.,  {Pinto} C.,  {Walton} D.~J.,  {Alston} W.~N.,  {Bachetti} M.,
  et~al., 2020, \mnras, 491, 5172

\bibitem[\protect\citeauthoryear{{King}, {Davies}, {Ward}, {Fabbiano} \&
  {Elvis}}{{King} et~al.}{2001}]{King2001}
{King} A.~R.,  {Davies} M.~B.,  {Ward} M.~J.,  {Fabbiano} G.,    {Elvis} M.,
  2001, \apjl, 552, L109

\bibitem[\protect\citeauthoryear{{Kobayashi}, {Ohsuga}, {Takahashi},
  {Kawashima}, {Asahina}, {Takeuchi} \& {Mineshige}}{{Kobayashi}
  et~al.}{2018}]{Kobayashi2018}
{Kobayashi} H.,  {Ohsuga} K.,  {Takahashi} H.~R.,  {Kawashima} T.,  {Asahina}
  Y.,  {Takeuchi} S.,    {Mineshige} S.,  2018, \pasj, 70, 22

\bibitem[\protect\citeauthoryear{{Kosec}, {Pinto}, {Fabian} \&
  {Walton}}{{Kosec} et~al.}{2018a}]{Kosec2018a}
{Kosec} P.,  {Pinto} C.,  {Fabian} A.~C.,    {Walton} D.~J.,  {2018a}, \mnras,
  473, 5680

\bibitem[\protect\citeauthoryear{{Kosec}, {Pinto}, {Walton} et~al.,}{{Kosec}
  et~al.}{2018b}]{Kosec2018b}
{Kosec} P.,  {Pinto} C.,  {Walton} D.~J.,    et~al., {2018b}, \mnras, 479, 3978

\bibitem[\protect\citeauthoryear{{Krolik}, {McKee} \& {Tarter}}{{Krolik}
  et~al.}{1981}]{Krolik1981}
{Krolik} J.~H.,  {McKee} C.~F.,    {Tarter} C.~B.,  1981, \apj, 249, 422

\bibitem[\protect\citeauthoryear{{Liu}, {Bregman}, {Bai}, {Justham} \&
  {Crowther}}{{Liu} et~al.}{2013}]{Liu2013}
{Liu} J.-F.,  {Bregman} J.~N.,  {Bai} Y.,  {Justham} S.,    {Crowther} P.,
  2013, \nat, 503, 500

\bibitem[\protect\citeauthoryear{{Lodders} \& {Palme}}{{Lodders} \&
  {Palme}}{2009}]{Lodders2009}
{Lodders} K.,  {Palme} H.,  2009, Meteoritics and Planetary Science Supplement,
  72, 5154

\bibitem[\protect\citeauthoryear{{Lumb}, {Warwick}, {Page} \& {De Luca}}{{Lumb}
  et~al.}{2002}]{Lumb2002}
{Lumb} D.~H.,  {Warwick} R.~S.,  {Page} M.,    {De Luca} A.,  2002, \aap, 389,
  93

\bibitem[\protect\citeauthoryear{{Mezcua}, {Civano}, {Fabbiano}, {Miyaji} \&
  {Marchesi}}{{Mezcua} et~al.}{2016}]{Mezcua2016}
{Mezcua} M.,  {Civano} F.,  {Fabbiano} G.,  {Miyaji} T.,    {Marchesi} S.,
  2016, \apj, 817, 20

\bibitem[\protect\citeauthoryear{{Middleton}, {Brightman}, {Pintore}
  et~al.,}{{Middleton} et~al.}{2019}]{Middleton2019}
{Middleton} M.~J.,  {Brightman} M.,  {Pintore} F.,    et~al., 2019, \mnras,
  486, 2

\bibitem[\protect\citeauthoryear{{Middleton}, {Heil}, {Pintore}, {Walton} \&
  {Roberts}}{{Middleton} et~al.}{2015a}]{Middleton2015a}
{Middleton} M.~J.,  {Heil} L.,  {Pintore} F.,  {Walton} D.~J.,    {Roberts}
  T.~P.,  2015a, \mnras, 447, 3243

\bibitem[\protect\citeauthoryear{{Middleton}, {Miller-Jones}, {Markoff},
  {Fender}, {Henze} et~al.,}{{Middleton} et~al.}{2013}]{Middleton2013}
{Middleton} M.~J.,  {Miller-Jones} J.~C.~A.,  {Markoff} S.,  {Fender} R.,
  {Henze} M.,    et~al., 2013, \nat, 493, 187

\bibitem[\protect\citeauthoryear{{Middleton}, {Walton}, {Fabian}, {Roberts},
  {Heil}, {Pinto} et~al.,}{{Middleton} et~al.}{2015b}]{Middleton2015b}
{Middleton} M.~J.,  {Walton} D.~J.,  {Fabian} A.,  {Roberts} T.~P.,  {Heil} L.,
   {Pinto} C.,    et~al., 2015b, \mnras, 454, 3134

\bibitem[\protect\citeauthoryear{{Middleton}, {Walton}, {Roberts} \&
  {Heil}}{{Middleton} et~al.}{2014}]{Middleton2014}
{Middleton} M.~J.,  {Walton} D.~J.,  {Roberts} T.~P.,    {Heil} L.,  2014,
  \mnras, 438, L51

\bibitem[\protect\citeauthoryear{{Miller}, {Walton}, {King}, {Reynolds},
  {Fabian} et~al.,}{{Miller} et~al.}{2013}]{Miller2013}
{Miller} J.~M.,  {Walton} D.~J.,  {King} A.~L.,  {Reynolds} M.~T.,  {Fabian}
  A.~C.,    et~al., 2013, \apjl, 776, L36

\bibitem[\protect\citeauthoryear{{Miller}, {Turner} \& {Reeves}}{{Miller}
  et~al.}{2008}]{Miller2008pca}
{Miller} L.,  {Turner} T.~J.,    {Reeves} J.~N.,  2008, \aap, 483, 437

\bibitem[\protect\citeauthoryear{{Miller}, {Turner}, {Reeves}, {George},
  {Kraemer} \& {Wingert}}{{Miller} et~al.}{2007}]{Miller2007pca}
{Miller} L.,  {Turner} T.~J.,  {Reeves} J.~N.,  {George} I.~M.,  {Kraemer}
  S.~B.,    {Wingert} B.,  2007, \aap, 463, 131

\bibitem[\protect\citeauthoryear{{Mittaz}, {Penston} \& {Snijders}}{{Mittaz}
  et~al.}{1990}]{Mittaz1990}
{Mittaz} J.~P.~D.,  {Penston} M.~V.,    {Snijders} M.~A.~J.,  1990, \mnras,
  242, 370

\bibitem[\protect\citeauthoryear{{Mizuno}, {Miyawaki}, {Ebisawa}, {Kubota},
  {Miyamoto} et~al.,}{{Mizuno} et~al.}{2007}]{Mizuno2007}
{Mizuno} T.,  {Miyawaki} R.,  {Ebisawa} K.,  {Kubota} A.,  {Miyamoto} M.,
  et~al., 2007, \pasj, 59, 257

\bibitem[\protect\citeauthoryear{{Motch}, {Pakull}, {Soria}, {Gris{\'e}} \&
  {Pietrzy{\'n}ski}}{{Motch} et~al.}{2014}]{Motch2014}
{Motch} C.,  {Pakull} M.~W.,  {Soria} R.,  {Gris{\'e}} F.,    {Pietrzy{\'n}ski}
  G.,  2014, \nat, 514, 198

\bibitem[\protect\citeauthoryear{{Mushtukov}, {Ingram}, {Middleton}, {Nagirner}
  \& {van der Klis}}{{Mushtukov} et~al.}{2019}]{Mushtukov2019a}
{Mushtukov} A.~A.,  {Ingram} A.,  {Middleton} M.,  {Nagirner} D.~I.,    {van
  der Klis} M.,  2019, \mnras, 484, 687

\bibitem[\protect\citeauthoryear{{Nandra}, {Barret}, {Barcons}, {Fabian}, {den
  Herder}, {Piro}, {Watson}, {Adami}, {Aird}, {Afonso} et~al.,}{{Nandra}
  et~al.}{2013}]{Nandra2013}
{Nandra} K.,  {Barret} D.,  {Barcons} X.,  {Fabian} A.,  {den Herder} J.-W.,
  {Piro} L.,  {Watson} M.,  {Adami} C.,  {Aird} J.,  {Afonso} J.~M.,    et~al.,
  2013, ArXiv e-prints

\bibitem[\protect\citeauthoryear{{Parker}, {Alston}, {Buisson}, {Fabian}
  et~al.,}{{Parker} et~al.}{2017b}]{Parker2017b}
{Parker} M.~L.,  {Alston} W.~N.,  {Buisson} D.~J.~K.,  {Fabian} A.~C.,
  et~al., 2017b, \mnras, 469, 1553

\bibitem[\protect\citeauthoryear{{Parker}, {Pinto}, {Fabian}, {Lohfink},
  {Buisson} et~al.,}{{Parker} et~al.}{2017a}]{Parker2017a}
{Parker} M.~L.,  {Pinto} C.,  {Fabian} A.~C.,  {Lohfink} A.,  {Buisson}
  D.~J.~K.,    et~al., 2017a, Nature, 543, 83

\bibitem[\protect\citeauthoryear{{Parker}, {Walton}, {Fabian} \&
  {Risaliti}}{{Parker} et~al.}{2014}]{Parker2014}
{Parker} M.~L.,  {Walton} D.~J.,  {Fabian} A.~C.,    {Risaliti} G.,  2014,
  \mnras, 441, 1817

\bibitem[\protect\citeauthoryear{{Pinto}, {Alston}, {Parker}, {Fabian}, {Gallo}
  et~al.,}{{Pinto} et~al.}{2018}]{Pinto2018a}
{Pinto} C.,  {Alston} W.,  {Parker} M.~L.,  {Fabian} A.~C.,  {Gallo} L.~C.,
  et~al., 2018, \mnras, 476, 1021

\bibitem[\protect\citeauthoryear{{Pinto}, {Alston}, {Soria}, {Middleton},
  {Walton} et~al.,}{{Pinto} et~al.}{2017}]{Pinto2017}
{Pinto} C.,  {Alston} W.,  {Soria} R.,  {Middleton} M.~J.,  {Walton} D.~J.,
  et~al., 2017, \mnras, 468, 2865

\bibitem[\protect\citeauthoryear{{Pinto}, {Kaastra}, {Costantini} \& {de
  Vries}}{{Pinto} et~al.}{2013}]{Pinto2013}
{Pinto} C.,  {Kaastra} J.~S.,  {Costantini} E.,    {de Vries} C.,  2013, A\&A,
  551, A25

\bibitem[\protect\citeauthoryear{Pinto, et al.}{2019}]{Pinto2019a} {Pinto} C., 
{Mehdipour} M.,  {Walton} D.~J.,  {Middleton} M.~J.,  {Roberts}
  T.~P.  et al., 2019, MNRAS.tmp, 3026


\bibitem[\protect\citeauthoryear{{Pinto}, {Middleton} \& {Fabian}}{{Pinto}
  et~al.}{2016}]{Pinto2016nature}
{Pinto} C.,  {Middleton} M.~J.,    {Fabian} A.~C.,  2016, Nature, 533, 64

\bibitem[\protect\citeauthoryear{{Pinto}, {Ness}, {Verbunt}, {Kaastra},
  {Costantini} \& {Detmers}}{{Pinto} et~al.}{2012}]{Pinto2012b}
{Pinto} C.,  {Ness} J.-U.,  {Verbunt} F.,  {Kaastra} J.~S.,  {Costantini} E.,
   {Detmers} R.~G.,  2012, \aap, 543, A134

\bibitem[\protect\citeauthoryear{{Pintore} \& {Zampieri}}{{Pintore} \&
  {Zampieri}}{2012}]{Pintore2012}
{Pintore} F.,  {Zampieri} L.,  2012, \mnras, 420, 1107

\bibitem[\protect\citeauthoryear{{Porquet} \& {Dubau}}{{Porquet} \&
  {Dubau}}{2000}]{Porquet2000}
{Porquet} D.,  {Dubau} J.,  2000, A\&AS, 143, 495

\bibitem[\protect\citeauthoryear{{Poutanen}, {Lipunova}, {Fabrika}, {Butkevich}
  \& {Abolmasov}}{{Poutanen} et~al.}{2007}]{Poutanen2007}
{Poutanen} J.,  {Lipunova} G.,  {Fabrika} S.,  {Butkevich} A.~G.,
  {Abolmasov} P.,  2007, \mnras, 377, 1187

\bibitem[\protect\citeauthoryear{{Prestwich}, {Tsantaki}, {Zezas}, {Jackson},
  {Roberts} et~al.,}{{Prestwich} et~al.}{2013}]{Prestwich2013}
{Prestwich} A.~H.,  {Tsantaki} M.,  {Zezas} A.,  {Jackson} F.,  {Roberts}
  T.~P.,    et~al., 2013, \apj, 769, 92

\bibitem[\protect\citeauthoryear{{Psaradaki}, {Costantini}, {Mehdipour} \&
  {D{\'{\i}}az Trigo}}{{Psaradaki} et~al.}{2018}]{Psaradaki2018}
{Psaradaki} I.,  {Costantini} E.,  {Mehdipour} M.,    {D{\'{\i}}az Trigo} M.,
  2018, \aap, 620, A129

\bibitem[\protect\citeauthoryear{{Rodriguez Castillo}, {Israel}
  et~al.,}{{Rodriguez Castillo} et~al.}{2019}]{RodriguezCastillo2019}
{Rodriguez Castillo} G.~A.,  {Israel} G.~L.,    et~al., 2019, arXiv e-prints,
  p. arXiv:1906.04791

\bibitem[\protect\citeauthoryear{{Sathyaprakash}, {Roberts} \&
  {Siwek}}{{Sathyaprakash} et~al.}{2019}]{Sathyaprakash2019b}
{Sathyaprakash} R.,  {Roberts} T.~P.,    {Siwek} M.~M.,  2019, \mnras, p.~1986

\bibitem[\protect\citeauthoryear{{Sathyaprakash}, {Roberts}, {Walton}, {Fuerst}
  et~al.,}{{Sathyaprakash} et~al.}{2019}]{Sathyaprakash2019a}
{Sathyaprakash} R.,  {Roberts} T.~P.,  {Walton} D.~J.,  {Fuerst} F.,    et~al.,
  2019, \mnras, p.~L104

\bibitem[\protect\citeauthoryear{{Shakura} \& {Sunyaev}}{{Shakura} \&
  {Sunyaev}}{1973}]{SS1973}
{Shakura} N.~I.,  {Sunyaev} R.~A.,  1973, \aap, 24, 337

\bibitem[\protect\citeauthoryear{{Soria}, {Wu} \& {Hunstead}}{{Soria}
  et~al.}{2000}]{Soria2000}
{Soria} R.,  {Wu} K.,    {Hunstead} R.~W.,  2000, \apj, 539, 445

\bibitem[\protect\citeauthoryear{{Stobbart}, {Roberts} \& {Wilms}}{{Stobbart}
  et~al.}{2006}]{Stobbart2006}
{Stobbart} A.-M.,  {Roberts} T.~P.,    {Wilms} J.,  2006, \mnras, 368, 397

\bibitem[\protect\citeauthoryear{{Sutton}, {Roberts} \& {Middleton}}{{Sutton}
  et~al.}{2013}]{Sutton2013}
{Sutton} A.~D.,  {Roberts} T.~P.,    {Middleton} M.~J.,  2013, \mnras, 435,
  1758

\bibitem[\protect\citeauthoryear{{Sutton}, {Roberts} \& {Middleton}}{{Sutton}
  et~al.}{2015}]{Sutton2015}
{Sutton} A.~D.,  {Roberts} T.~P.,    {Middleton} M.~J.,  2015, \apj, 814, 73

\bibitem[\protect\citeauthoryear{{Takeuchi}, {Ohsuga} \&
  {Mineshige}}{{Takeuchi} et~al.}{2013}]{Takeuchi2013}
{Takeuchi} S.,  {Ohsuga} K.,    {Mineshige} S.,  2013, \pasj, 65, 88

\bibitem[\protect\citeauthoryear{{Tarter}, {Tucker} \& {Salpeter}}{{Tarter}
  et~al.}{1969}]{Tarter1969}
{Tarter} C.~B.,  {Tucker} W.~H.,    {Salpeter} E.~E.,  1969, \apj, 156, 943

\bibitem[\protect\citeauthoryear{{Tsygankov}, {Mushtukov}, {Suleimanov} \&
  {Poutanen}}{{Tsygankov} et~al.}{2016}]{Tsygankov2016}
{Tsygankov} S.~S.,  {Mushtukov} A.~A.,  {Suleimanov} V.~F.,    {Poutanen} J.,
  2016, \mnras, 457, 1101

\bibitem[\protect\citeauthoryear{{Tully}, {Courtois}, {Dolphin}, {Fisher},
  {H{\'e}raudeau} et~al.,}{{Tully} et~al.}{2013}]{Tully2013}
{Tully} R.~B.,  {Courtois} H.~M.,  {Dolphin} A.~E.,  {Fisher} J.~R.,
  {H{\'e}raudeau} P.,    et~al., 2013, \aj, 146, 86

\bibitem[\protect\citeauthoryear{{Turner}, {Abbey}, {Arnaud}, {Balasini},
  {Barbera} et~al.,}{{Turner} et~al.}{2001}]{Turner2001}
{Turner} M.~J.~L.,  {Abbey} A.,  {Arnaud} M.,  {Balasini} M.,  {Barbera} M.,
  et~al., 2001, A\&A, 365, L27

\bibitem[\protect\citeauthoryear{{Urquhart} \& {Soria}}{{Urquhart} \&
  {Soria}}{2016}]{Urquhart2016}
{Urquhart} R.,  {Soria} R.,  2016, \mnras, 456, 1859

\bibitem[\protect\citeauthoryear{{van Peet}, {Costantini}, {M{\'e}ndez},
  {Paerels} \& {Cottam}}{{van Peet} et~al.}{2009}]{vanPeet2009}
{van Peet} J.~C.~A.,  {Costantini} E.,  {M{\'e}ndez} M.,  {Paerels} F.~B.~S.,
   {Cottam} J.,  2009, \aap, 497, 805

\bibitem[\protect\citeauthoryear{{Vaughan} \& {Fabian}}{{Vaughan} \&
  {Fabian}}{2004}]{Vaughan2004}
{Vaughan} S.,  {Fabian} A.~C.,  2004, \mnras, 348, 1415

\bibitem[\protect\citeauthoryear{{Walton}, {F{\"u}rst}, {Bachetti}, {Barret},
  {Brightman} et~al.,}{{Walton} et~al.}{2016b}]{Walton2016b}
{Walton} D.~J.,  {F{\"u}rst} F.,  {Bachetti} M.,  {Barret} D.,  {Brightman} M.,
     et~al., 2016b, \apjl, 827, L13

\bibitem[\protect\citeauthoryear{{Walton}, {F{\"u}rst}, {Heida}, {Harrison},
  {Barret} et~al.,}{{Walton} et~al.}{2018}]{Walton2018a}
{Walton} D.~J.,  {F{\"u}rst} F.,  {Heida} M.,  {Harrison} F.~A.,  {Barret} D.,
    et~al., 2018, \apj, 856, 128

\bibitem[\protect\citeauthoryear{{Walton}, {Middleton}, {Pinto}, {Fabian},
  {Bachetti} et~al.,}{{Walton} et~al.}{2016a}]{Walton2016a}
{Walton} D.~J.,  {Middleton} M.~J.,  {Pinto} C.,  {Fabian} A.~C.,  {Bachetti}
  M.,    et~al., 2016a, \apjl, 826, L26

\bibitem[\protect\citeauthoryear{{Walton}, {Tomsick}, {Madsen}, {Grinberg},
  {Barret} et~al.,}{{Walton} et~al.}{2016c}]{Walton2016c}
{Walton} D.~J.,  {Tomsick} J.~A.,  {Madsen} K.~K.,  {Grinberg} V.,  {Barret}
  D.,    et~al., 2016c, \apj, 826, 87

\bibitem[\protect\citeauthoryear{{Walton}, {Pinto}, {Nowak}, {Bachetti},
  {Sathyaprakash} et~al.,}{{Walton} et~al.}{2019}]{Walton2020}
{Walton} D.~J.,  {Pinto} C.,  {Nowak} M.,  {Bachetti} M.,  {Sathyaprakash} R.,
    et~al., 2019, submitted to MNRAS, arXiv:1911.09622

\bibitem[\protect\citeauthoryear{Wang, Soria \& Wang}{2019}]{Wang2019} Wang C., Soria R., Wang J., 2019, ApJ, 883, 44

\bibitem[\protect\citeauthoryear{{Webb}, {Cseh}, {Lenc}, {Godet}, {Barret}
  et~al.,}{{Webb} et~al.}{2012}]{Webb2012}
{Webb} N.,  {Cseh} D.,  {Lenc} E.,  {Godet} O.,  {Barret} D.,    et~al., 2012,
  Science, 337, 554

\bibitem[\protect\citeauthoryear{{Weng} \& {Feng}}{{Weng} \&
  {Feng}}{2018}]{Weng2018}
{Weng} S.-S.,  {Feng} H.,  2018, \apj, 853, 115

\bibitem[\protect\citeauthoryear{{Winter}, {Mushotzky} \& {Reynolds}}{{Winter}
  et~al.}{2007}]{Winter2007}
{Winter} L.~M.,  {Mushotzky} R.~F.,    {Reynolds} C.~S.,  2007, \apj, 655, 163

\end{thebibliography}


\label{lastpage}

\end{document}